\documentclass[pre]{revtex4-1}

\usepackage{latexsym} \usepackage{graphicx}
\usepackage{amssymb} \usepackage{bm}
\usepackage{color}\usepackage{amsmath}
\usepackage{subfigure}

\begin{document}

\title{Mixing and demixing of binary mixtures of polar chiral active particles \footnote{Electronic supplementary information (ESI) available}}
%Collective Behavior of %Structure Formation
\author{Bao-quan Ai$^{1}$}\email[Email: ]{aibq@scnu.edu.cn}
\author{Zhi-gang Shao$^{1}$}\email[Email: ]{zgshao@scnu.edu.cn}
\author{Wei-rong Zhong$^{2}$}\email[Email: ]{wrzhong@jnu.edu.cn}
\affiliation{$^{1}$Guangdong Provincial Key Laboratory of Quantum Engineering and Quantum Materials, School of Physics and Telecommunication
Engineering, South China Normal University, Guangzhou 510006, China.}
\affiliation{$^{2}$Siyuan Laboratory, Guangzhou Key Laboratory of Vacuum Coating Technologies and
New Energy Materials, Department of Physics, Jinan University, Guangzhou 510632, China.}

\date{\today}

\begin{abstract}
\indent We study a binary mixture of polar chiral (counterclockwise or clockwise) active particles in a two-dimensional box with periodic boundary conditions. Beside the excluded volume interactions between particles, particles are also subject to the polar velocity alignment. From the extensive Brownian dynamics simulations, it is found that the particle configuration (mixing or demixing) is determined by the competition between the chirality difference and the polar velocity alignment. When the chirality difference competes with the polar velocity alignment, the clockwise particles aggregate in one cluster and the counterclockwise particles aggregate in the other cluster, thus particles are demixed and can be separated. However, when the chirality difference or the polar velocity alignment is dominated, particles are mixed. Our findings could be used for the experimental pursuit of the separation of binary mixtures of chiral active particles.
\end{abstract}

\maketitle

\section{Introduction}
\indent Active matter in biological and physical systems has been studied theoretically and experimentally \cite{Hanggi,Bechinger,Marchetti0,Reichhardt0,Cates,Reichhardt1x}. Unlike passive particles, active particles, also known as self-propelled particles or microswimmers and nanoswimmers, are capable of taking up energy from their environment and converting it into directed motion. Understanding the collective behavior of systems composed of the active matter can provide insight into out-of-equilibrium phenomena associated with many biological systems, such as fish schools, bacterial suspensions, and artificial microswimmers.

\indent Demixing and separation strategies of active particles based on their swimming properties \cite{Volpe,McCandlish,Stenhammar,Ma,Smrek,Harder,Maggi, Yang, Costanzo,Berdakin,Nourhani,Weber,Kumari,Mijalkov,Reichhardt,Scholz,Nguyen,Dolai,Agrawal,Wysocki,Ai1,Ai2,Ai3,Shin} are of utmost importance for various branches of science and engineering. Usually, there are three types of mixed particles. 1)Mixture of active and passive particles. McCandlish and coworkers\cite{McCandlish} realized the spontaneous segregation of active and passive particles, freely swimming in a two-dimensional box. Stenhammar and coworkers \cite{Stenhammar} investigated the phase behavior and kinetics of a monodisperse mixture of active and passive particles and found that motility of the active component can trigger phase separation. In the mixture of passive particles and eccentric self-propelled active particles\cite{Ma}, the eccentric particles can push passive particles to form a large dense dynamic cluster when the eccentricity of active particles are high enough. Smrek and Kremer found that small activity differences drive phase separation in active-passive polymer mixtures\cite{Smrek}. 2)Mixture of active particles with different properties. Swimming cells with different motilities can be separated by using the centrifugation \cite{Maggi}. Separation of binary mixtures of colloids is realized by using self-driven artificial microswimmers \cite{Yang}. Costanzo and coworkers \cite{Costanzo} proposed a separation method for run-and-tumble particles in terms of their motility in a microchannel. Asymmetric obstacles can also induce the separation of self-propelled particles with different motilities \cite{Berdakin}. Weber and coworkers \cite{Weber} studied the exclusive role of diffusivity differences in binary mixtures of equal-sized particles and found that differences in diffusion constant alone suffice to drive phase separation in binary mixtures. Demixing of binary mixtures of particles with different effective diffusion coefficients is significantly enhanced in the presence of an external potential\cite{Kumari}. 3)Mixture of chiral active particles.  The class of chiral active matter includes a variety of biological circle swimmers, such as E. coli, which swim circularly when close to walls and interfaces\cite{Leonardo,DiLuzio}, as well as magnetotactic bacteria in rotating external fields\cite{Cebers} and  sperm cells\cite{Riedel}.  Mijalkov and Volpe\cite{Mijalkov} found that chiral microswimmers can be sorted on the basis of their swimming properties by employing some simple static patterns in their environment. Particles with the same radius but different chirality move in different directions in asymmetrically patterned arrays\cite{Reichhardt}. In the circle confinement clockwise and counter-clockwise rotating robots move collectively and phase separate via spinodal decomposition\cite{Scholz}. In the previous work\cite{Ai2}, we proposed a sorting method, where the mixed chiral particles can be separated when the angular speed of the obstacles, the angular speed of active particles, and the self-propulsion speed satisfy a certain relation.

\indent  Most studies on demixing and separation of active particles have considered only the excluded volume interactions between particles. The effect of alignment interactions between particles on demixing and separation has not yet been investigated. In this paper, we numerically study the binary mixture of chiral active particles in the presence of the excluded volume interactions as well as the polar velocity alignment. It is found that when the chirality difference competes with the polar velocity alignment, the clockwise particles aggregate in one cluster and the counterclockwise particles aggregate in the other cluster, thus particles are demixed and can be separated. In the previous works, chiral particle separation was realized by using the external driving, the separation channel, or the special obstacles. However, in the present work, the internal competition between the chirality difference and the polar velocity alignment can induce particle separation.

\section{Model and methods}
%\indent Recently, there is a strong interest in a class of self-propelled particles which change their direction of motion autonomously.  Therefore, a deeper understanding of  chiral sorting mechanisms is of great fundamental importance.

\indent We consider the binary system of  chiral active particles ($N/2$ counterclockwise and $N/2$ clockwise particles) of radius $r$ in a two-dimensional box of size $L\times L$ with periodic boundary conditions. Besides the excluded volume interactions, we also consider the polar velocity alignment interactions between particles \cite{Vicsek}. The dynamics of each particle is described by the position $\mathbf{r}_i\equiv (x_i,y_i)$ of its center and the orientation $\theta_i$ of the polar axis $\mathbf{n}_i\equiv(\cos\theta_i,\sin\theta_i)$. The orientation $\theta_i$ is determined by the rotation diffusion, the constant torque acting on the particles (which is responsible for circular swimming), and the mutual interaction between neighboring swimmers. The translational and rotational diffusion are assumed to be not coupled and the translational diffusion is assumed to be negligibly small, thus we do not consider the translational diffusion in the model. We studied this model by solving the overdamped Langiven equations (thus neglecting hydrodynamic interactions between particles)\cite{Levis,Liechen,Mart¨ªn-Gomez}
\begin{equation}\label{eq1}
\frac{d\mathbf{r}_i}{dt}=v_0 \mathbf{n}_i+\mu \sum_{j=1}^{N}\mathbf{F}_{ij},
\end{equation}
\begin{equation}\label{eq2}
\frac{d\theta_i}{dt}=\Omega_i+\frac{g}{\pi R^2}\sum_{j\in\partial_i}  \sin(\theta_j-\theta_i)+\sqrt{2D_r}\xi_i(t),
\end{equation}
where $v_0$ is the self-propulsion speed and $\mu$ is the mobility. $D_r$ is the rotational diffusion coefficient and $\xi_i(t)$ is unit-variance Gaussian white noise with zero mean. The angular velocity $\Omega_i=\pm \omega$ (particles experience a constant torque leading to circular motion) and its sign determines the chirality of the particle. The positive $\omega$ denotes the chirality difference. We define particles as the counterclockwise particles for positive $\Omega_i$ and the clockwise particles for negative $\Omega_i$. For the convenience of discussion, we focus on equimolar mixtures, where the number of the counterclockwise particles is equal to the number of the clockwise particles.

\indent The interactions between particles are taken as short-ranged harmonic repulsive force: $\mathbf{F}_{ij}=k(2r-r_{ij})\mathbf{\hat{r}}_{ij}$ if particles overlap ($r_{ij}<2r$), and $\mathbf{F}_{ij}=0$ otherwise. $r_{ij}=|\mathbf{r}_i-\mathbf{r}_j|$ is the distance between Brownian particle $i$ and particle $j$ and $\mathbf{\hat{r}}_{ij}=(\mathbf{r}_i-\mathbf{r}_j)/r_{ij}$. Here $k$ describes the spring constant. In order to mimic hard particles, we use a larger value for the product of spring constant and mobility $\mu k(=100)$, thus ensuring that particle overlaps decay quickly. We tested that the presented results are robust against reasonable changes in $\mu k$. The polar velocity alignment interactions are introduced as a torque in Eq. (\ref{eq2}), and their strength is controlled by the coupling constant $g\geq 0$. The sum in Eq. (\ref{eq2}) runs over neighbor with a radius $R$ around particle $i$.

\indent To quantify the spatial distribution of the two particle types, we divide the system in $M$ square sub-regions of area $(L\times L)/M$ and the segregation coefficient $S$\cite{Yang1} is given by
\begin{equation}\label{}
  S=\frac{1}{N}\sum_{i=1}^{M}|N_i^{CW}-N_i^{CCW}|,
\end{equation}
where $N_i^{CW}$ ($N_i^{CCW}$) is number of clockwise (counterclockwise) particles in the $i$-th sub-region. With this definition, $S\to 0$ for uniform distribution of clockwise and counterclockwise particles, and $S\to 1$ for complete segregation.

\indent The orientational order can be characterized by the polar order parameter $P$,
\begin{equation}\label{order}
  P=\langle\left|\frac{1}{N}\sum_{j=1}^{N}e^{i\theta_j(t)}\right|\rangle,
\end{equation}
where $\langle...\rangle$ denotes time average and $N$ stands for the total number of particles in the system. When $P\to 1$ all particles move in the same direction and when $P\to 0$ particles move in any direction with equal probability. We define
the ratio between the area occupied by particles and the total available area as the packing fraction $\rho=N\pi r^2/(L\times L)$.

\indent  In order to describe the characteristic cluster size of the single particle specie in the binary mixtures,  we define the relative radial distribution function as the autocorrelation function \cite{Scholz,Nguyen},
\begin{equation}\label{g1}
  g_{\alpha\beta}(\mathbf{r}_1,\mathbf{r}_2)=\langle \rho_{\alpha}(\mathbf{r}_1)\rho_{\alpha}(\mathbf{r}_2) \rangle+\langle \rho_{\beta}(\mathbf{r}_1)\rho_{\beta}(\mathbf{r}_2) \rangle-\langle \rho_{\alpha}(\mathbf{r}_1)\rho_{\beta}(\mathbf{r}_2) \rangle-\langle \rho_{\beta}(\mathbf{r}_1)\rho_{\alpha}(\mathbf{r}_2) \rangle,
\end{equation}
here $\rho_{I}(\mathbf{r})=\sum_{i=1}^{N_{I}}\delta(\mathbf{r}-\mathbf{r}_i)$ is the particle density for the particle specie $\mu$, where $I=\alpha$, $\beta$. In the homogeneous, isotropic system, Eq. (\ref{g1}) reduces to $g_{\alpha\beta}(r)$ with $r=\left|\mathbf{r}_1-\mathbf{r}_2\right|$. The characteristic cluster size is determined by the first non-trivial zero of $g_{\alpha\beta}(r)$ \cite{Scholz,Nguyen}.

\indent Times and lengths are in units of the elastic time $\frac{1}{\mu k}$ and the particle radius $r$.  The parameters in the dimensionless forms can be rewritten as $\hat{v}_0=\frac{v_0}{\mu k r}$, $\hat{\omega}=\frac{\omega}{\mu k}$, $\hat{D}_r=\frac{D_r}{\mu k}$, and $\hat{g}=\frac{g}{\pi R^2 \mu k}$. From now on, we will use only the dimensionless variables and shall omit the hat for all quantities occurring in the above equations.  We explore the behaviors of the system by varying the self-propulsion speed $v_0$, the rotational diffusion rate $D_r$, the angular velocity $\omega$, and the alignment interaction strength $g$.
\section{Results and discussion}
\indent In our simulations, particle positions are initialized with a uniform random distribution inside the box, and orientations are random over the interval $[0, 2\pi]$. Eqs. (\ref{eq1}) and (\ref{eq2}) are integrated numerically using a Runge-Kutta algorithm. The integration step time was chosen to be smaller than $10^{-3}$ and the total integration time was more than $2\times10^5$ (this time is sufficient to ensure that the density profile of the system has reached steady state shown in Figure S1 in Electronic supplementary information (ESI)). We have considered $10^2$ realizations to improve accuracy and minimize statistical errors. Our numerical calculations were done by use of a Fortran CUDA environment implemented on a modern desktop GPU. This scheme allowed for a speed-up of a factor of the order $10^3$ times as compared to a common present-day CPU method. Unless otherwise noted, our simulations are under the parameter sets $L=40.0$, $M=10\times 10=100$, and $N=1024 (\rho=0.50)$.

\begin{figure}
\centering
\subfigure[$g=0.0$]{\label{fig:subfig:a}
\includegraphics[width=0.35\linewidth]{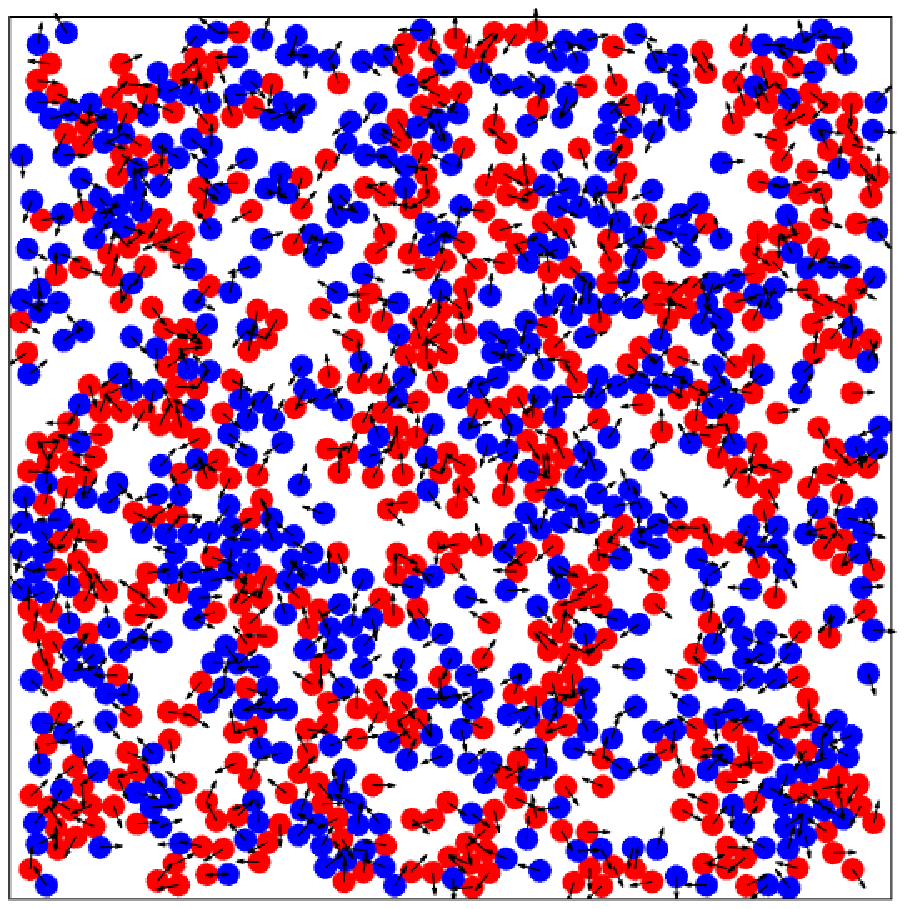}}
\hspace{0.1\linewidth}
\subfigure[$g=0.01$]{\label{fig:subfig:b}
\includegraphics[width=0.35\linewidth]{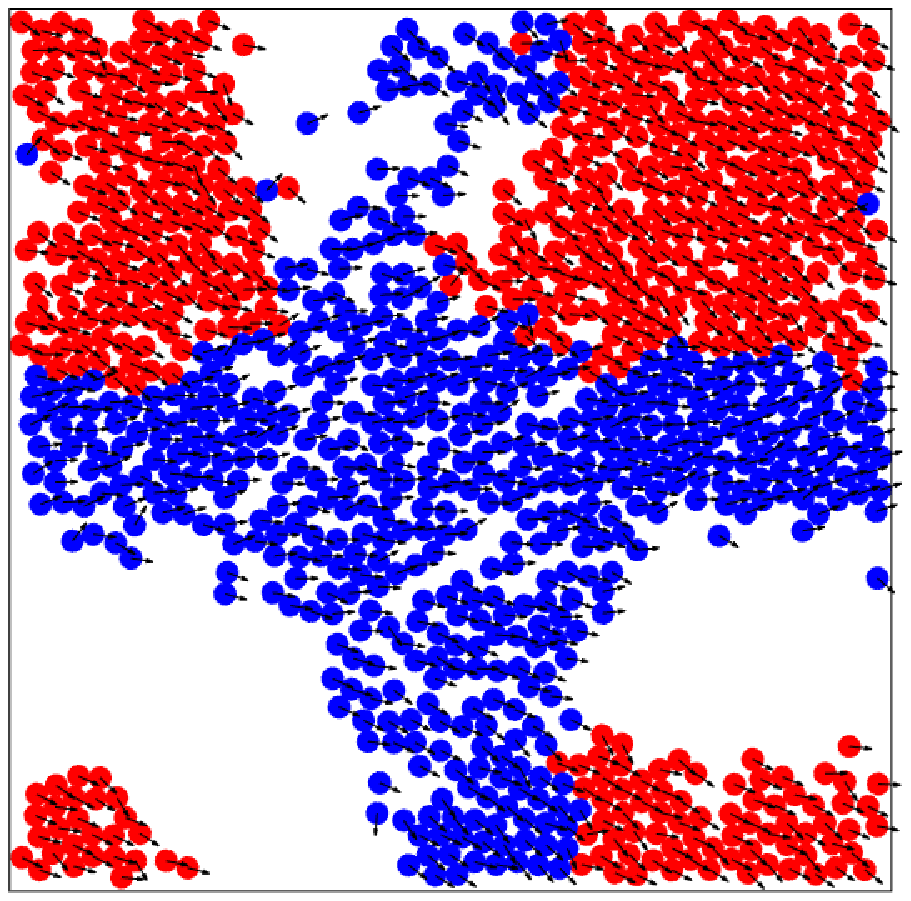}}
\vfill\vspace{0.01\linewidth}
\subfigure[$g=0.1$]{\label{fig:subfig:a}
\includegraphics[width=0.35\linewidth]{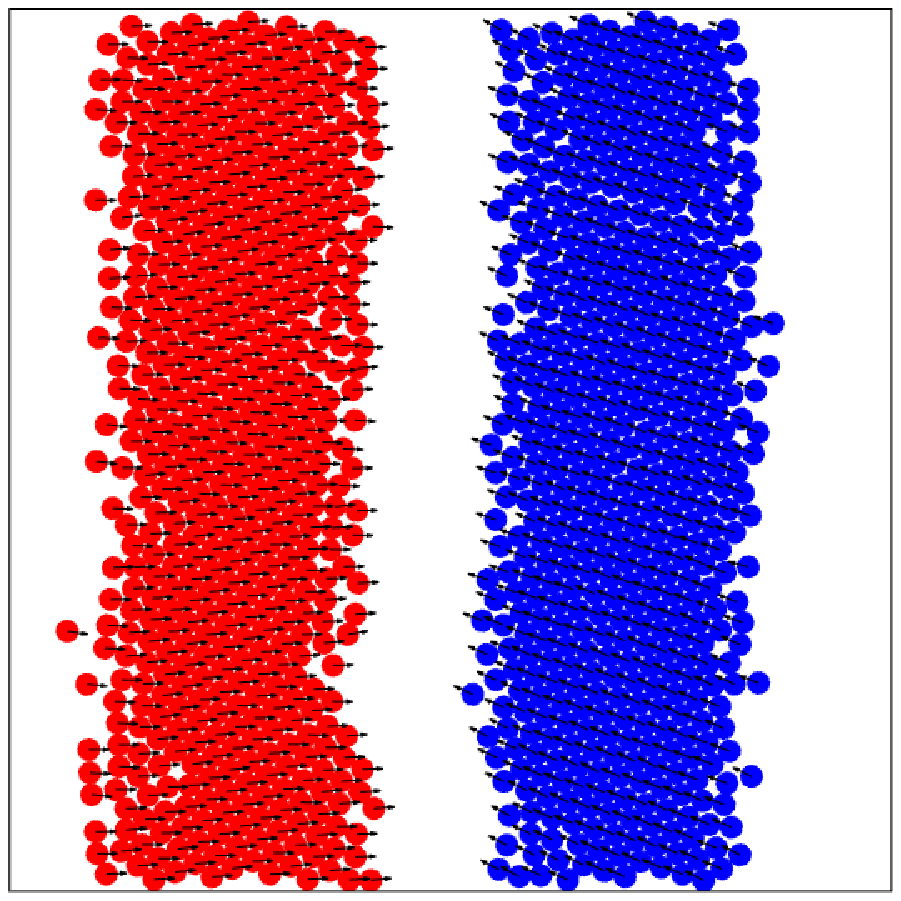}}
\hspace{0.1\linewidth}
\subfigure[$g=0.3$]{\label{fig:subfig:b}
\includegraphics[width=0.35\linewidth]{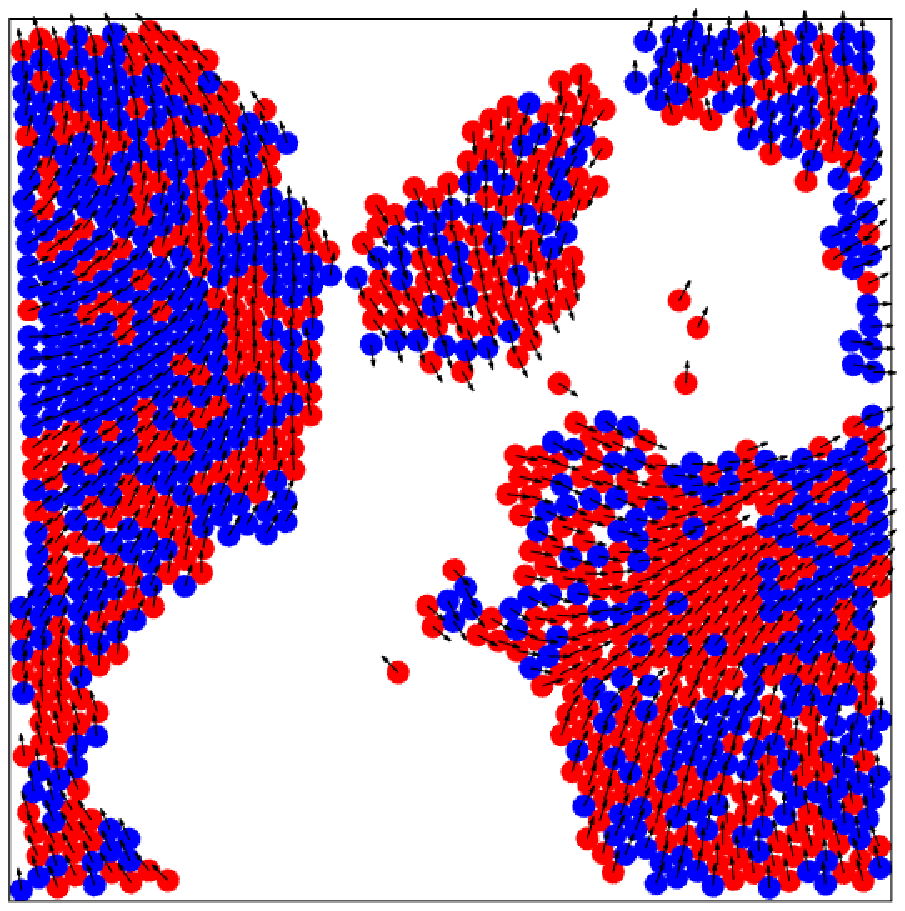}}
\caption{Snapshots of a mixture of the clockwise (blue) and counterclockwise (red) particles for different values of $g$ at $\omega=0.5$. (a)$g=0.0$. (b)$g=0.01$. (c)$g=0.1$. (d)$g=0.3$. The other parameters are $\rho=0.50$, $v_0=1.0$, and $D_r=0.001$.}
\label{fig:subfig}
\end{figure}

\indent For binary mixtures of chiral active particles, the change of the self-propulsion orientation $\theta$ is determined by the parameters $\omega$, $g$, and $D_r$. The angular velocity $\omega$ determines the chirality difference, particles are exactly the same at $\omega=0.0$. Particles are randomly segregated for small $g$ and aggregated for large $g$. The rotational diffusion coefficient $D_r$ describes the fluctuation of the angular velocity. When $D_r$ is fixed, the particle configuration (mixing or demixing) is determined by the competition between $\omega$ and $g$.

\indent Figure 1 shows the snapshots of mixed chiral active particles for different values of $g$ at $\omega=0.5$.  The chirality difference separates the particle group and the alignment interaction promotes particle aggregation.
For zero alignment interaction ($g=0$)(shown in Fig. 1(a)), particles do circular motions with small circular radius ($R=v_0/\omega=2$ is far less than the length $L$), the same kind of particle randomly aggregates in very small clusters. In the presence of the periodic boundary conditions, the size of small clusters do not increases with time, thus particles are mixed on the whole (see Movie 1,Electronic supplementary information (ESI)).  Note that if the confinement is applied, the size of small clusters gradually increases with time and finally symmetric demixing patterns appear\cite{Scholz}. When $g$ increases, the alignment interactions take effect, on the one hand the alignment interaction promotes particle aggregation, but on the other hand the chirality difference separates the particle group. The competition between the chirality difference and the alignment interaction determines the particle configuration (mixing or demixing). When the chirality difference ($\omega=0.5$) competes with the polar velocity alignment ($g=0.1$)(shown in Fig. 1(c)), the clockwise particles aggregate in one cluster and the counterclockwise particles aggregate in the other cluster, moreover particles in the same cluster move together. In this case, one cluster moves clockwise and the other cluster moves counterclockwise, due to the existence of the excluded volume interaction, these two clusters are mutually exclusive, therefore two types of particles are demixed (see Movie 2, Electronic supplementary information (ESI)) . On further increasing $g$ (shown in Fig. 1(d)), the aggregation gradually dominates the transport and the chirality difference can be neglected for large $g$ (e. g. $g=0.3$). In this case, particles tend to randomly aggregate in one cluster and move together, so particles are completely mixed (see Movie 3, Electronic supplementary information (ESI)).

\begin{figure}
\centering
\subfigure[$\omega=0.0$]{\label{fig:subfig:a}
\includegraphics[width=0.35\linewidth]{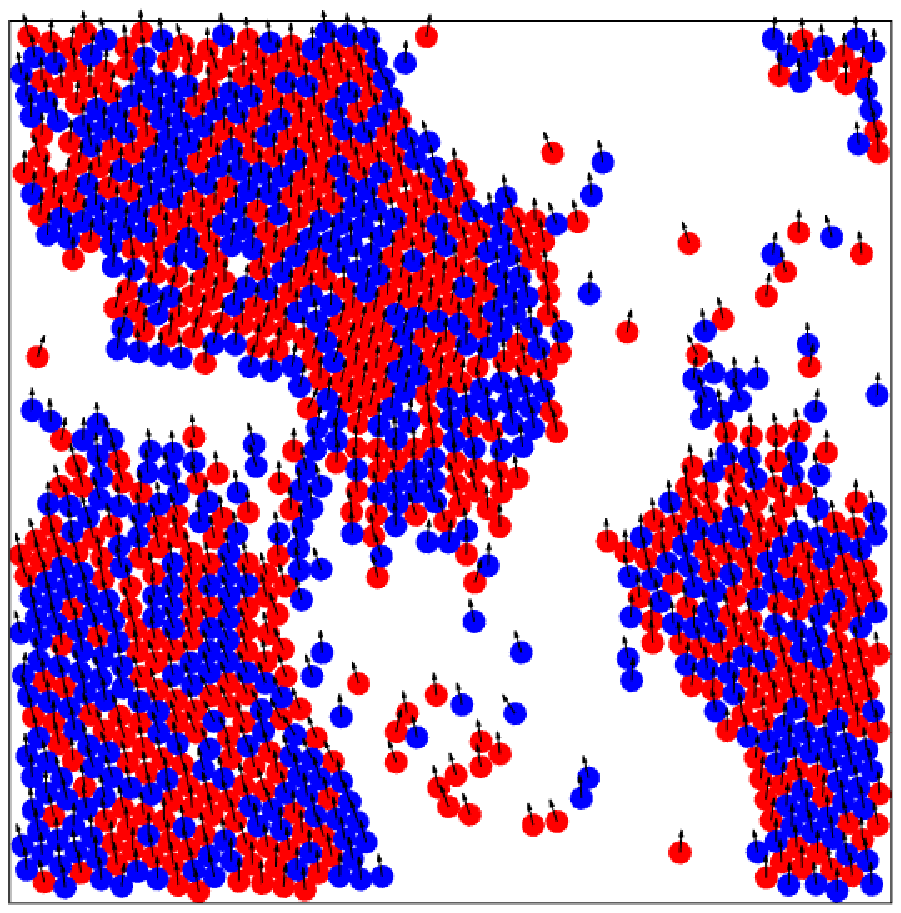}}
\hspace{0.1\linewidth}
\subfigure[$\omega=0.3$]{\label{fig:subfig:b}
\includegraphics[width=0.35\linewidth]{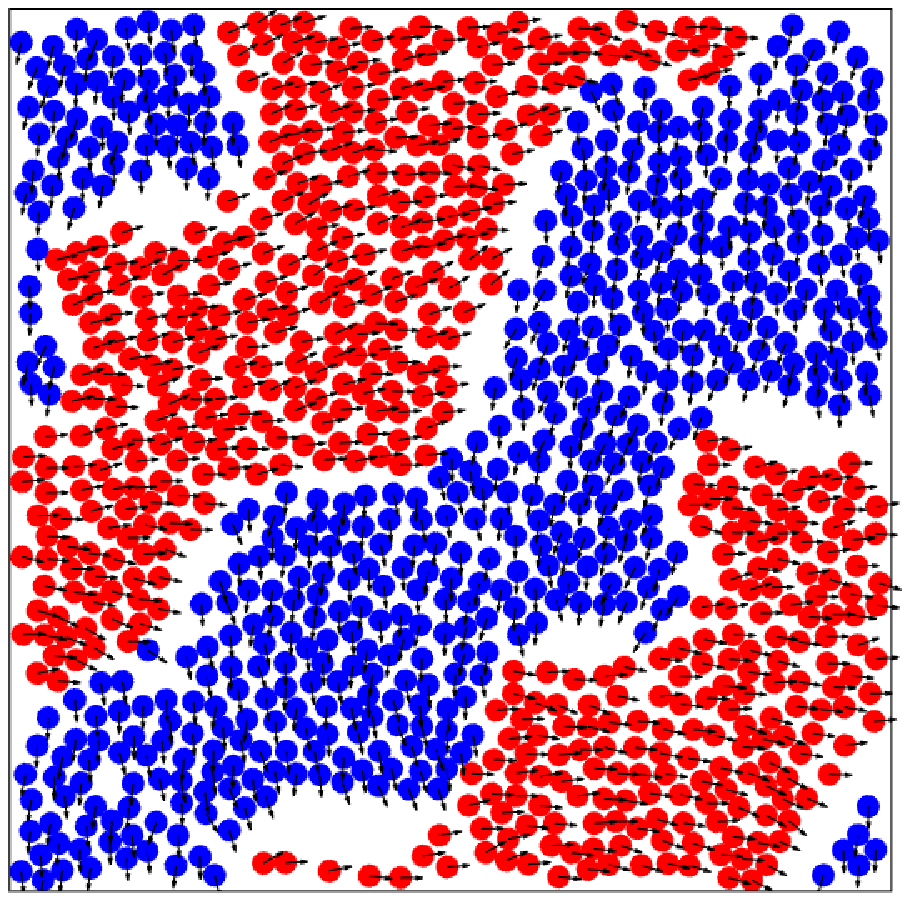}}
\vfill\vspace{0.01\linewidth}
\subfigure[$\omega=1.5$]{\label{fig:subfig:a}
\includegraphics[width=0.35\linewidth]{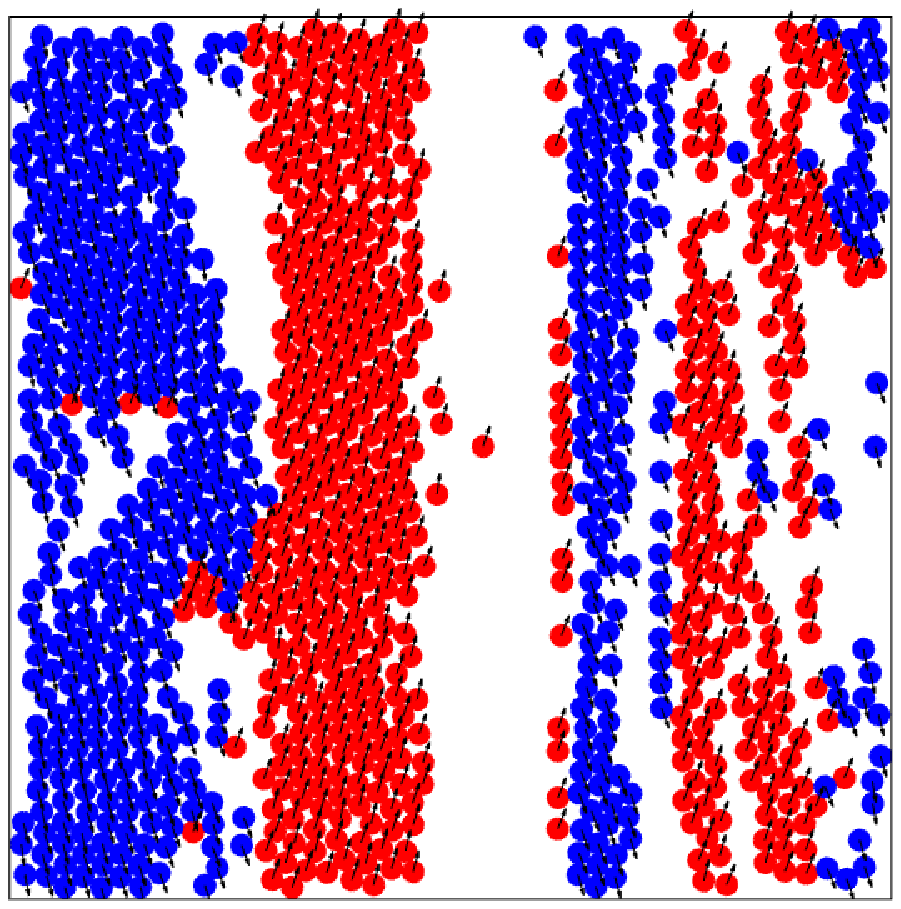}}
\hspace{0.1\linewidth}
\subfigure[$\omega=3.0$]{\label{fig:subfig:b}
\includegraphics[width=0.35\linewidth]{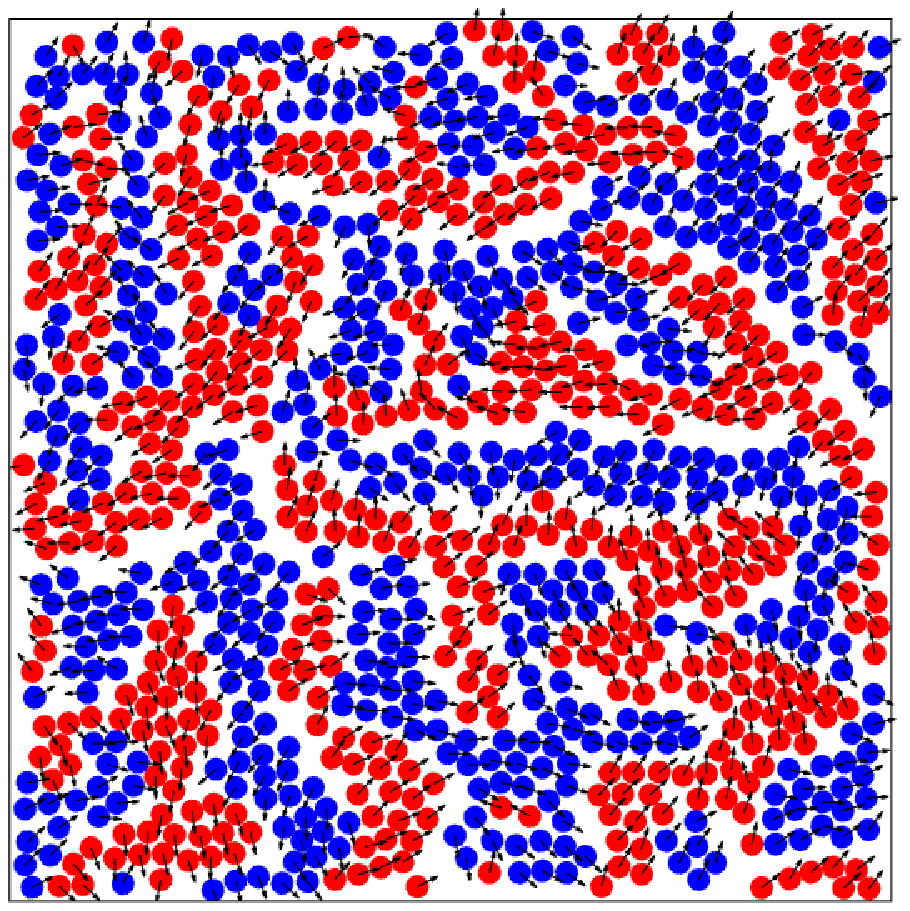}}
\caption{Snapshots of a mixture of the clockwise (blue) and counterclockwise (red) particles for different values of $\omega$ at $g=0.1$. (a)$\omega=0.0$. (b)$\omega=0.3$. (c)$\omega=1.5$. (d)$\omega=3.0$. The other parameters are $\rho=0.50$, $v_0=1.0$, and $D_r=0.001$.}
\label{fig:subfig}
\end{figure}

\indent Figure 2 describes the snapshots of mixed chiral active particles for different values of $\omega$ at $g=0.1$. When $\omega=0.0$, the polar velocity alignment completely dominates the transport, particles randomly aggregate in one cluster and move in the same direction shown in Fig. 2(a). Due to the zero chirality difference, all particles are identical, so there is no particle separation.  On increasing $\omega$, the chirality difference increases, particles with different chirality will gradually aggregate in different clusters.  When the chirality difference competes with the alignment interaction (e. g. $\omega=0.3$ and $1.5$), particles with the same chirality aggregate in the same cluster and move together, therefore two kinds of particles are demixed (shown in Figs. 2(b) and 2(c)). However, for large values of $\omega$ (e.g. $\omega=3.0$), the radius of circular motion ($\propto 1/\omega$) is small, particles move in many small regions. In each small region, two kinds of particles are demixed. However, on the whole many small clusters appear and particles are mixed (shown in Fig. 2(d)).

\begin{figure}[htbp]
\begin{center}
\vspace{0.1\linewidth}
\includegraphics[width=0.7\linewidth]{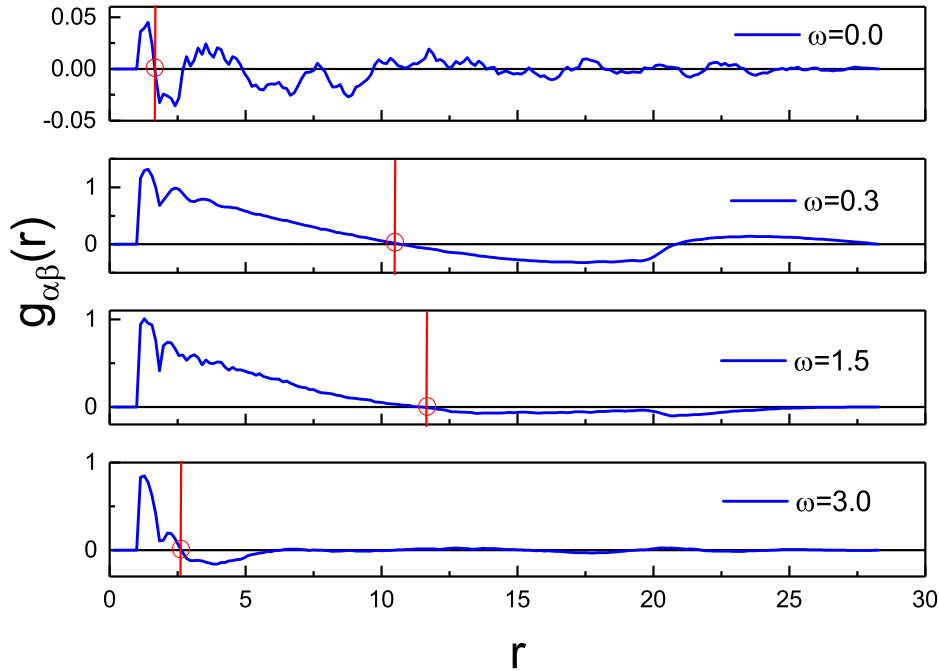}
\caption{Relative radial distribution function $g_{\alpha\beta}(r)$ for different values of $\omega$ at $t=2\times 10^5$.  The first non-trivial root (marked by
circles) is a quantifier for the cluster size of the single particle specie. The other parameter are $g=0.1$, $\rho=0.50$, $v_0=1.0$, and $D_r=0.001$.}
\end{center}
\end{figure}
\indent In order to describe the characteristic cluster size of the single particle specie in the binary mixtures (shown in Fig. 2), we have plotted the radial dependence of $g_{\alpha\beta}(r)$ in Fig. 3 for different values of $\omega$ at $g=0.1$.  The cluster size of the single particle specie is determined by the first non-trivial zero of $g_{\alpha\beta}(r)$(marked by circles). It is found that the cluster size of the single particle specie is large (e. g. $\omega=0.3$, $1.5$) when the chirality difference competes with the polar velocity alignment. The cluster size is small when the chirality difference(e. g. $\omega=3.0$) or the polar velocity alignment (e. g. $\omega=0$) dominates the dynamics.  The large cluster size of the single particle specie indicates the particle demixing.

\indent From the above figures, we can conclude that particles are demixed and can be separated when the chirality difference competes with the polar velocity alignment. In order to describe the particle configuration (mixing or demixing), we present the segregation coefficient $S$ and the corresponding polar order parameter $P$ as functions of $g$, $\omega$, $D_r$, and $v_0$ in Figs. 4-7. Note that every curve in these figures is obtained from the statistical average of 100 independent simulations and error bars in the curves are calculated based on 100 numerical runs.

\begin{figure}
\centering
\subfigure{\label{fig:subfig:a}
\includegraphics[width=0.4\linewidth]{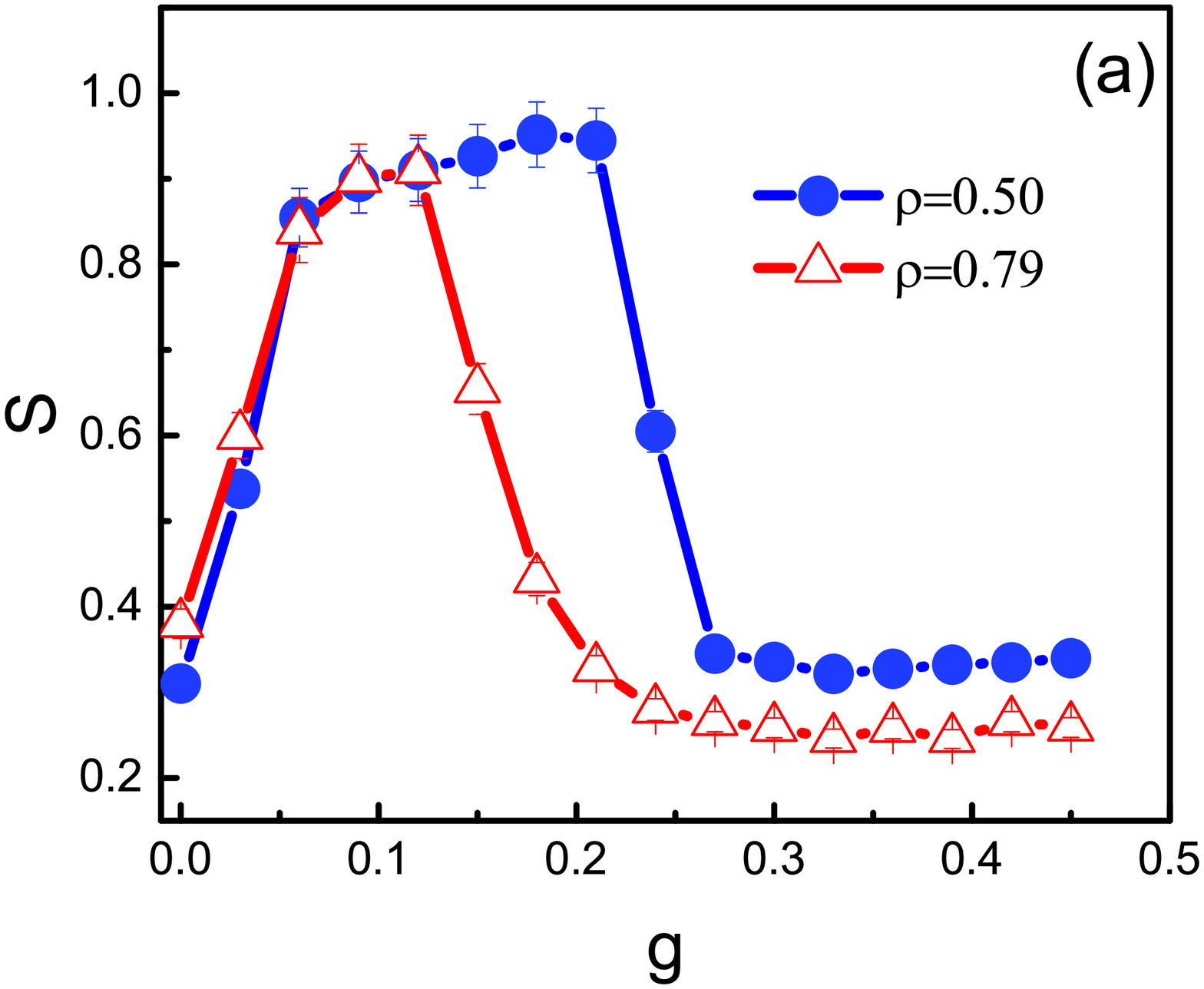}}
\hspace{0.1\linewidth}
\subfigure{\label{fig:subfig:b}
\includegraphics[width=0.4\linewidth]{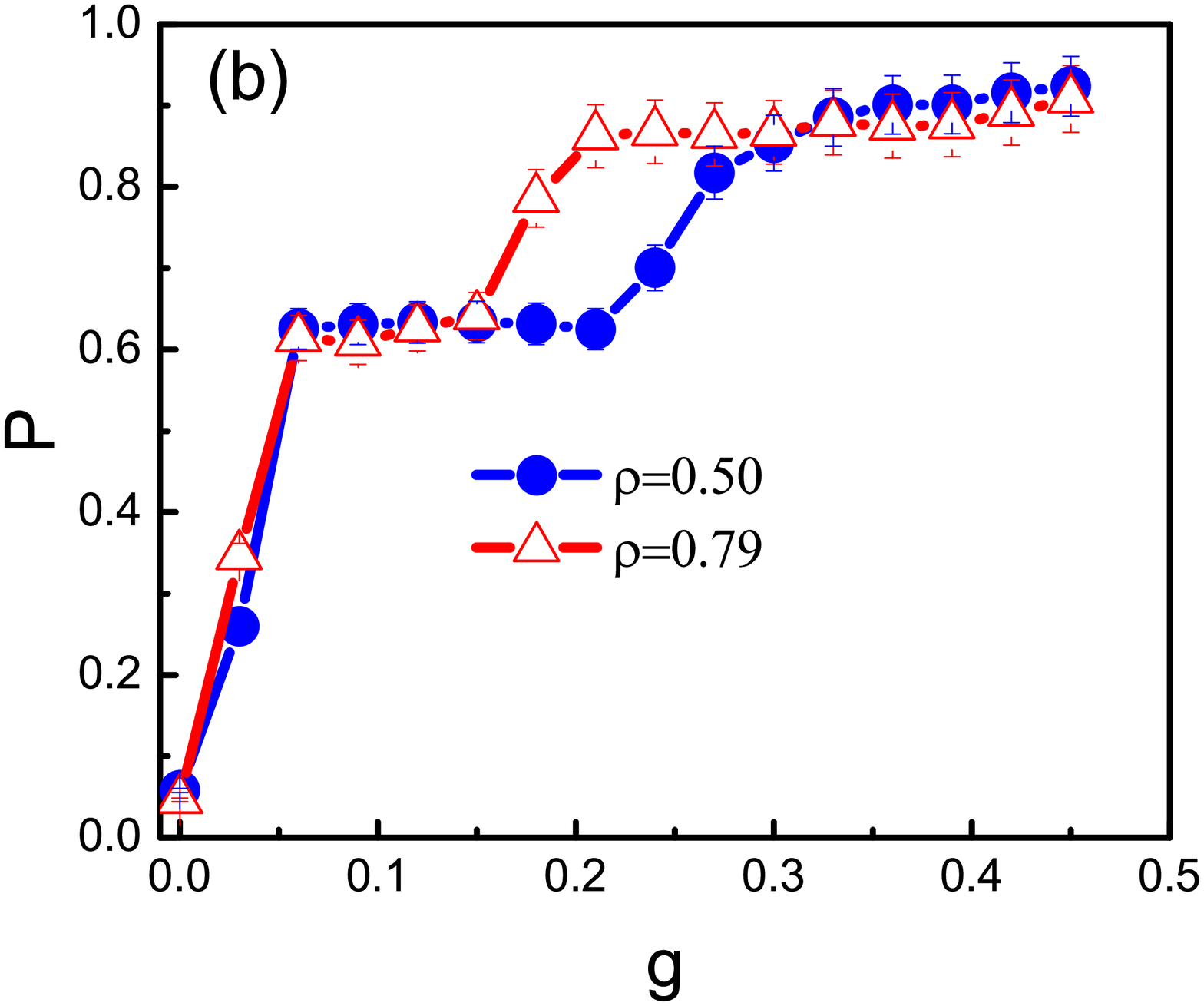}}
\vfill\vspace{0.001\linewidth}
\subfigure{\label{fig:subfig:a}
\includegraphics[width=0.4\linewidth]{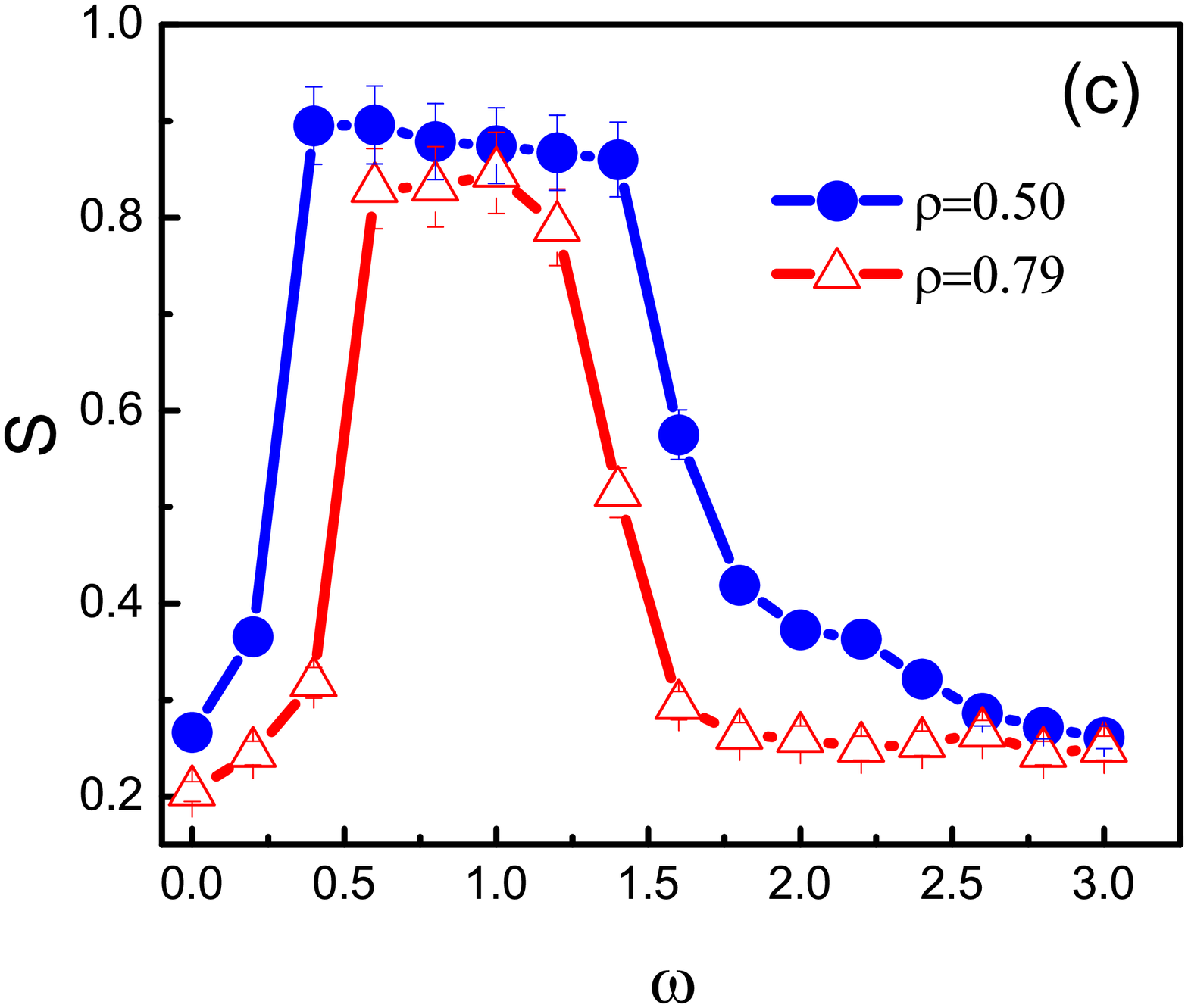}}
\hspace{0.1\linewidth}
\subfigure{\label{fig:subfig:b}
\includegraphics[width=0.4\linewidth]{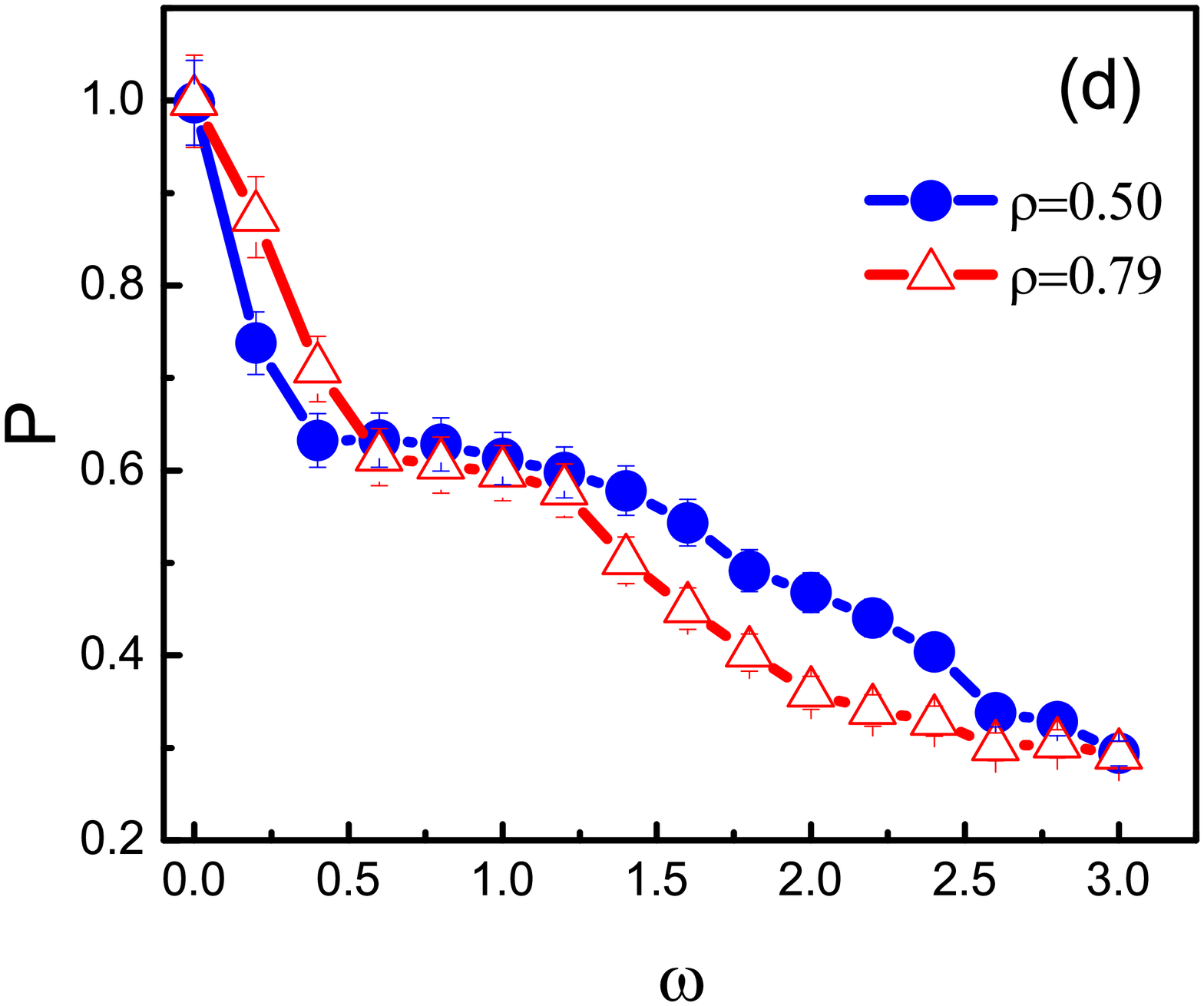}}
\caption{(a)Segregation coefficient $S$ as a function of the alignment interaction intensity $g$ at $\omega=0.5$. (b)Polar order parameter $P$ as a function of the alignment interaction intensity $g$ at $\omega=0.5$.
(c)Segregation coefficient $S$ as a function of the angular velocity $\omega$ at $g=0.1$. (d)Polar order parameter $P$ as a function of the angular velocity $\omega$ at $g=0.1$. The error bars represent the standard error of the mean.  The other parameters are $v_0=1.0$ and $D_r=0.001$. }
\label{fig:subfig}
\end{figure}

\indent Figure 4(a) describes the segregation coefficient $S$ as functions of the alignment interaction intensity $g$ at $\omega=0.5$. When $g\to 0$, particles are randomly segregated and mixed, thus the segregation coefficient is small. When $g>0.3$, the alignment interaction between particles dominates the dynamics, particles are aggregated and the chirality difference can be neglected, thus particles are mixed ($S$ is small). There exists a region of $g$ (the chirality difference competes with the polar velocity alignment) where $S>0.8$, which indicates that two types of particles are demixed. The corresponding polar order parameter $P$ is illustrated in Fig. 4(b). When $g$ increases from zero to $0.5$, the aggregate effect gradually becomes important, so the polar order parameter $P$ increases from $0$ to $1$. Note that in the curve there exists a step ($P\simeq 0.6$) which corresponds to particle demixing.

The segregation coefficient $S$ is illustrated in Fig. 4(c) as a function of $\omega$ at $g=0.1$.  When $\omega\to 0$, all particles are identical and the polar velocity alignment dominates the transport, particles are mixed ($S$ is small).  When $\omega$ is large (e. g. $>2.0$), the chirality difference dominates the transport, many small clusters appear, $S$ is small. There exists a region of $\omega$ where two types of particles are demixed and can be separated ($S>0.8$). On increasing $\omega$ form zero, the chirality difference gradually dominates the transport and the polar interaction becomes negligible, so the polar order parameter $P$ decreases from $1$ to $0$ (shown in Fig. 4(d)). Similar to Fig. 4(b), there is also a step (corresponding to particle demixing) in the curve of Fig. 4(d).

\indent Note that the parameter range ($g$ and $\omega$)corresponding to particle demixing reduces when the packing fraction increases from $0.50$ to $0.79$.  When the packing fraction increases, the average distance between particles becomes small, so the role of the polar alignment is more pronounced (see Figure S2 in Electronic supplementary information (ESI)). At the same time, the role of the excluded volume interactions between particles is also pronounced, which blocks the particle aggregation.

\begin{figure}[htbp]
\begin{center}
\vspace{0.1\linewidth}
\includegraphics[width=0.7\linewidth]{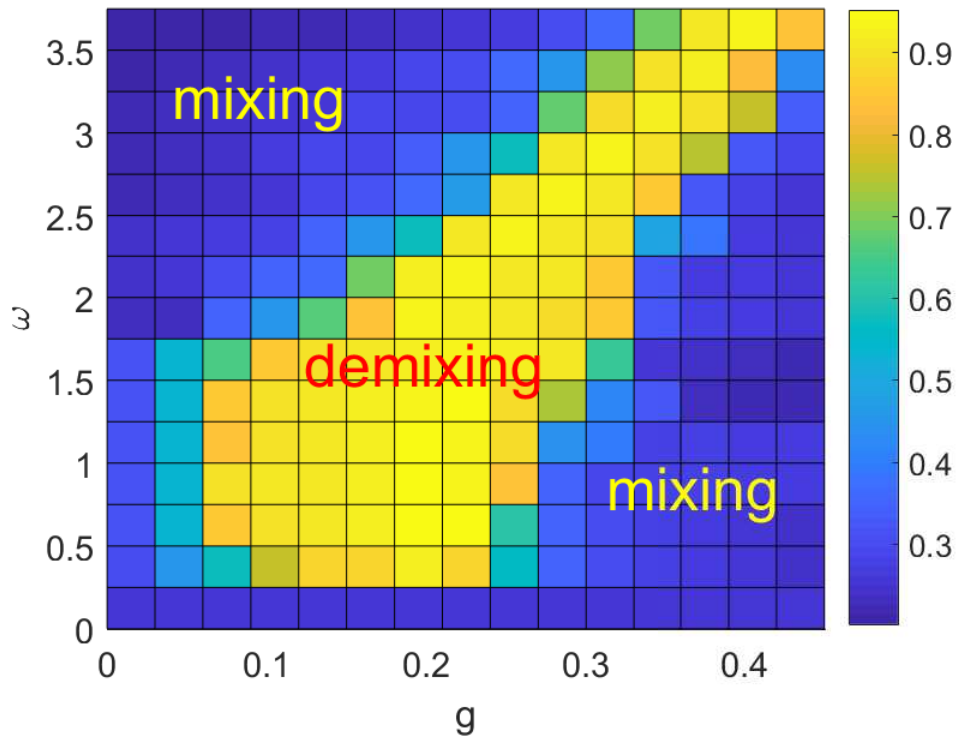}
\caption{Contour plots of the segregation coefficient $S$ as a function of the system parameters $g$ and $\omega$ at $\rho=0.50$, $v_0=1.0$, and $D_r=0.001$.}
\end{center}
\end{figure}

\indent To study in more detail the dependence of the segregation coefficient on $g$ and $\omega$, we plotted contour plots of the segregation coefficient $S$ as functions of $g$ and $\omega$ in Fig. 5. It is found that particles are demixed in the diagonal region and mixed in the nondiagonal region, which shows that particles are demixed when the chirality difference competes with the polar velocity alignment. The parameter range ($g$ and $\omega$)corresponding to particle demixing reduces when $g$ and $\omega$ increase.  Note that this phase diagram will change when the system parameters are varied.

\indent In order to further investigate the demxing of binary mixtures, the effects of the rotational diffusion coefficient $D_r$ and the self-propulsion speed $v_0$ on $S$ and $P$ are shown in Figs. 6 and 7.  We focus on three typical cases: 1) the chirality difference is dominated ($g=0.001$ and $\omega=0.5$),  2) the chirality difference competes with the polar velocity alignment ($g=0.1$ and $\omega=0.5$), and 3) the polar velocity alignment dominates the dynamics ($g=0.3$ and $\omega=0.5$).

\begin{figure}
\centering
\subfigure{\label{fig:subfig:a}
\includegraphics[width=0.4\linewidth]{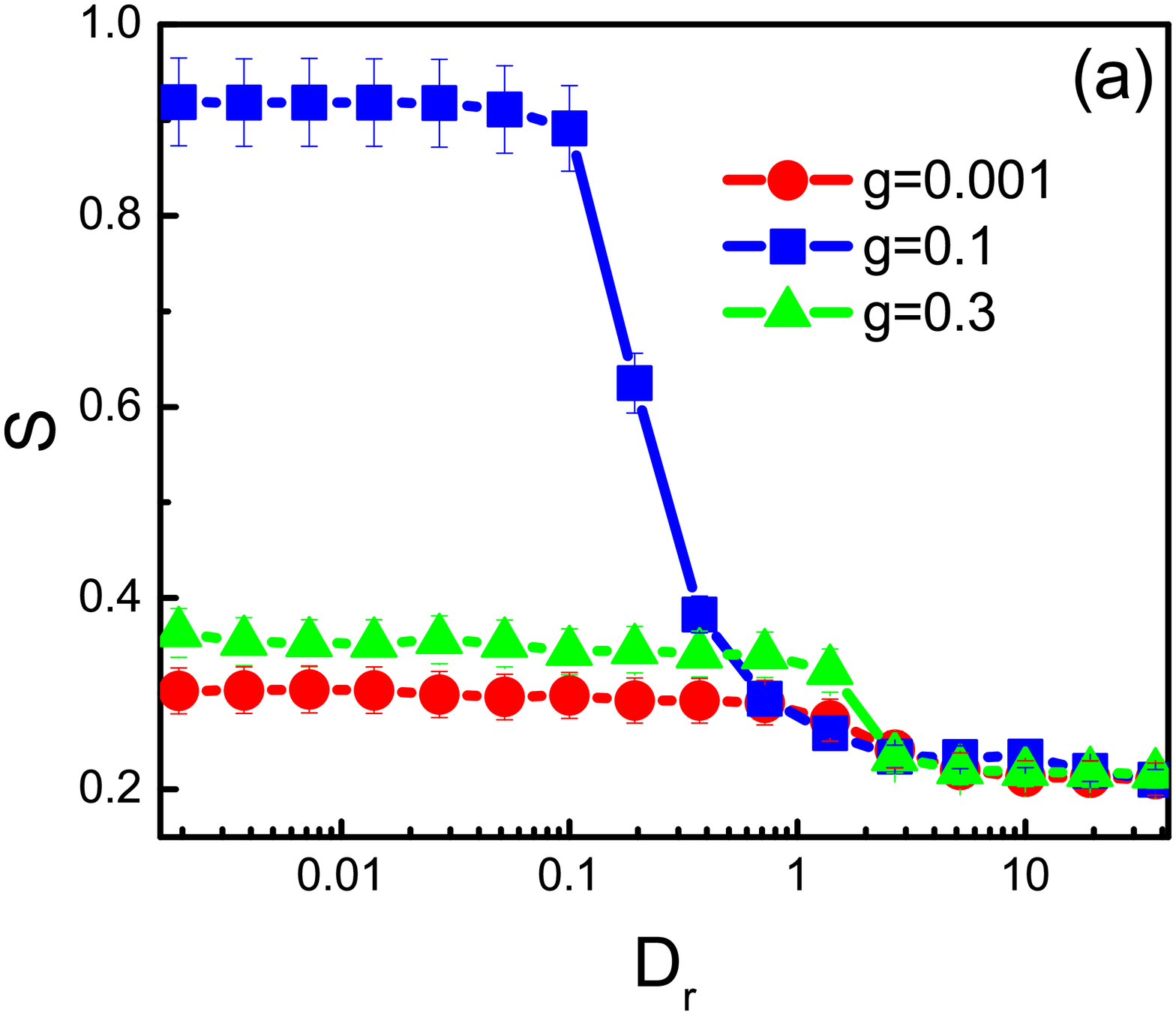}}
\hspace{0.1\linewidth}
\subfigure{\label{fig:subfig:b}
\includegraphics[width=0.4\linewidth]{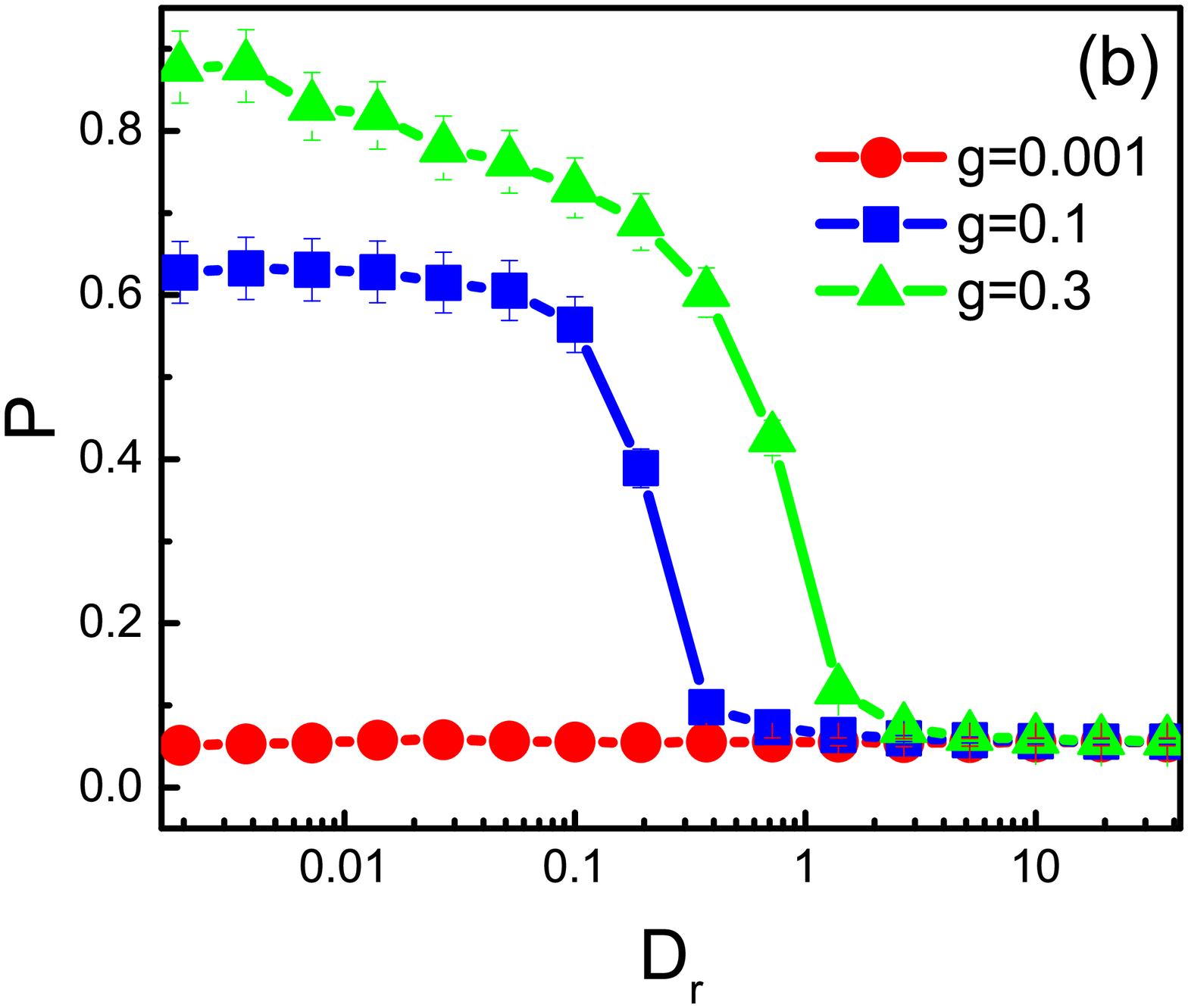}}
\caption{(a) Segregation coefficient $S$ as a function of the rotational diffusion coefficient $D_r$ for different values of $g$. (b)Polar order parameter $P$ as a function of the rotational diffusion coefficient $D_r$ for different values of $g$. The error bars represent the standard error of the mean. The other parameters are $\rho=0.50$, $v_0=1.0$, and $\omega=0.5$.}
\label{fig:subfig}
\end{figure}

\indent Figures 6(a) and 6(b) indicate the dependence of $S$ and $P$ on the rotational diffusion coefficient $D_r$ for three cases. When $D_r\to 0$, the rotational diffusion can be neglected, thus $S$ and $P$ is maximal. On increasing $D_r$, the rotational diffusion  gradually becomes important, both $S$ and $P$ decreases. When $D_r\to \infty$, the self-propelled angle changes very fast, the chirality and the alignment interaction becomes negligible, thus both $S$ and $P$ tend to zero. Therefore, both $S$ and $P$ decrease monotonously when the rotational diffusion coefficient increases. Note that for the case of $g=0.001$, the chirality of particle dominates the transport, the polar order parameter $P$ is very small and almost independent of $\omega$.

\begin{figure}
\centering
\subfigure{\label{fig:subfig:a}
\includegraphics[width=0.4\linewidth]{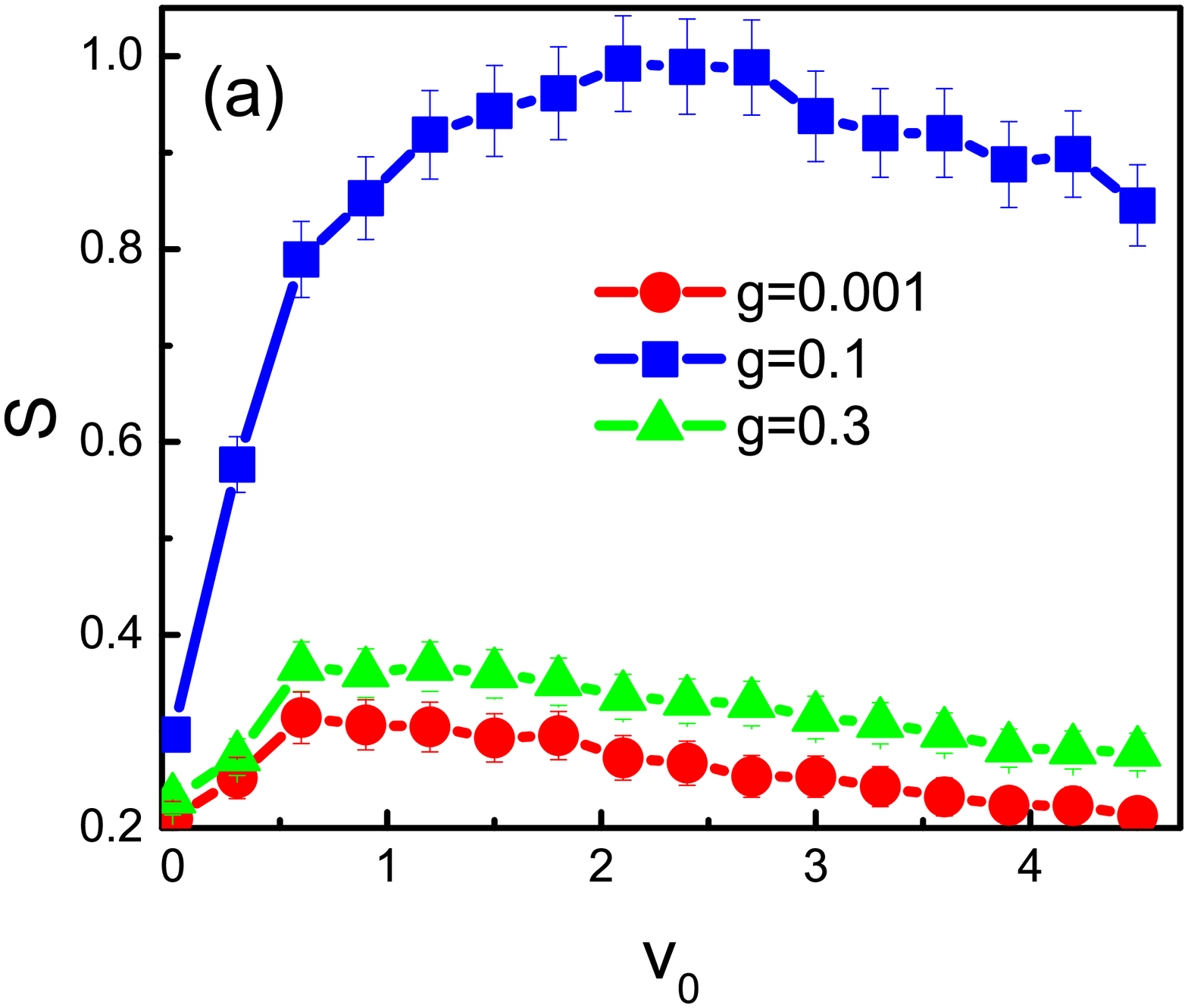}}
\hspace{0.1\linewidth}
\subfigure{\label{fig:subfig:b}
\includegraphics[width=0.4\linewidth]{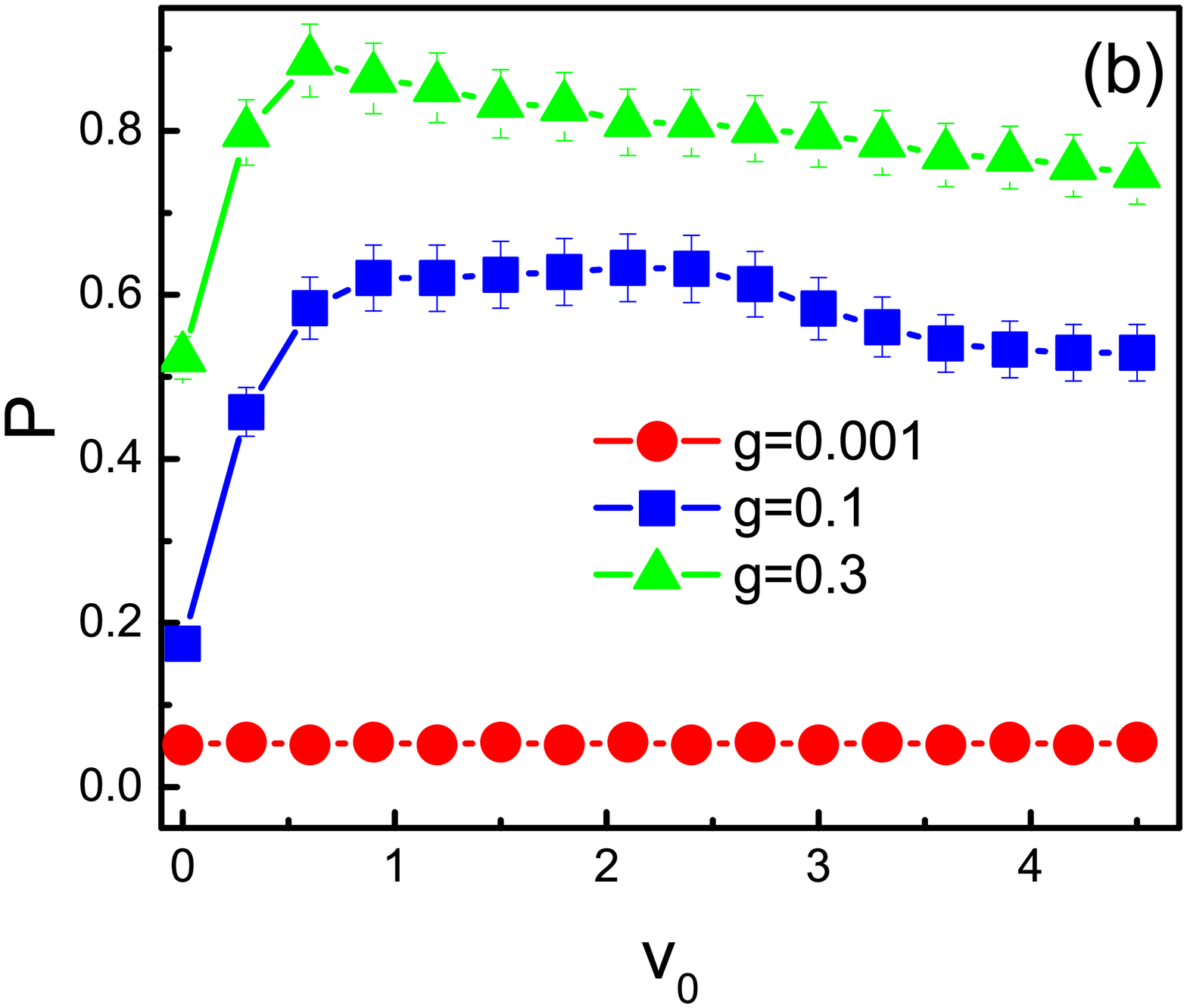}}
\caption{(a) Segregation coefficient $S$ as a function of the self-propulsion speed $v_0$ for different values of $g$. (b)Polar order parameter $P$ as a function of a function of the self-propulsion speed $v_0$ for different values of $g$. The error bars represent the standard error of the mean. The other parameters are $\rho=0.50$, $D_r=0.001$, and $\omega=0.5$.}
\label{fig:subfig}
\end{figure}

\indent In Figs. 7(a) and 7(b), $S$ and $P$ are depicted as a function of the self-propulsion speed $v_0$ for three cases. The circular motion radius of a single chiral particles is $R=\frac{v_0}{\omega}$. Particles with small circular motion radius are easier to aggregate than those with large circular motion radius. From Eq. (1), we can see that the dynamics of particles is determined by the self-propulsion and the excluded volume force. When $v_0\to 0$, the excluded volume force dominates the dynamics, the chirality difference becomes negligible, particles are randomly distributed, thus $S$ and $P$ are small. When $v_0$ is large, the self-propulsion dominates the dynamics and the excluded volume effect relatively reduces. In this case, particles do circular motions with large radius (which blocks the aggregation of particles) and tend to move in their own direction, so $S$ and $P$ decrease. Therefore, there exists an optimal value of $v_0$ at which $S$  and $P$ take their maximal values. Note that this is different from the achiral particle case where motility induced phase separation exists only for large speed\cite{Michael}. Similar to Fig. 6(b), the polar order parameter $P$ is very small and almost independent of $v_0$ for the case of $g=0.001$.

\begin{figure}
\centering
\subfigure{\label{fig:subfig:a}
\includegraphics[width=0.4\linewidth]{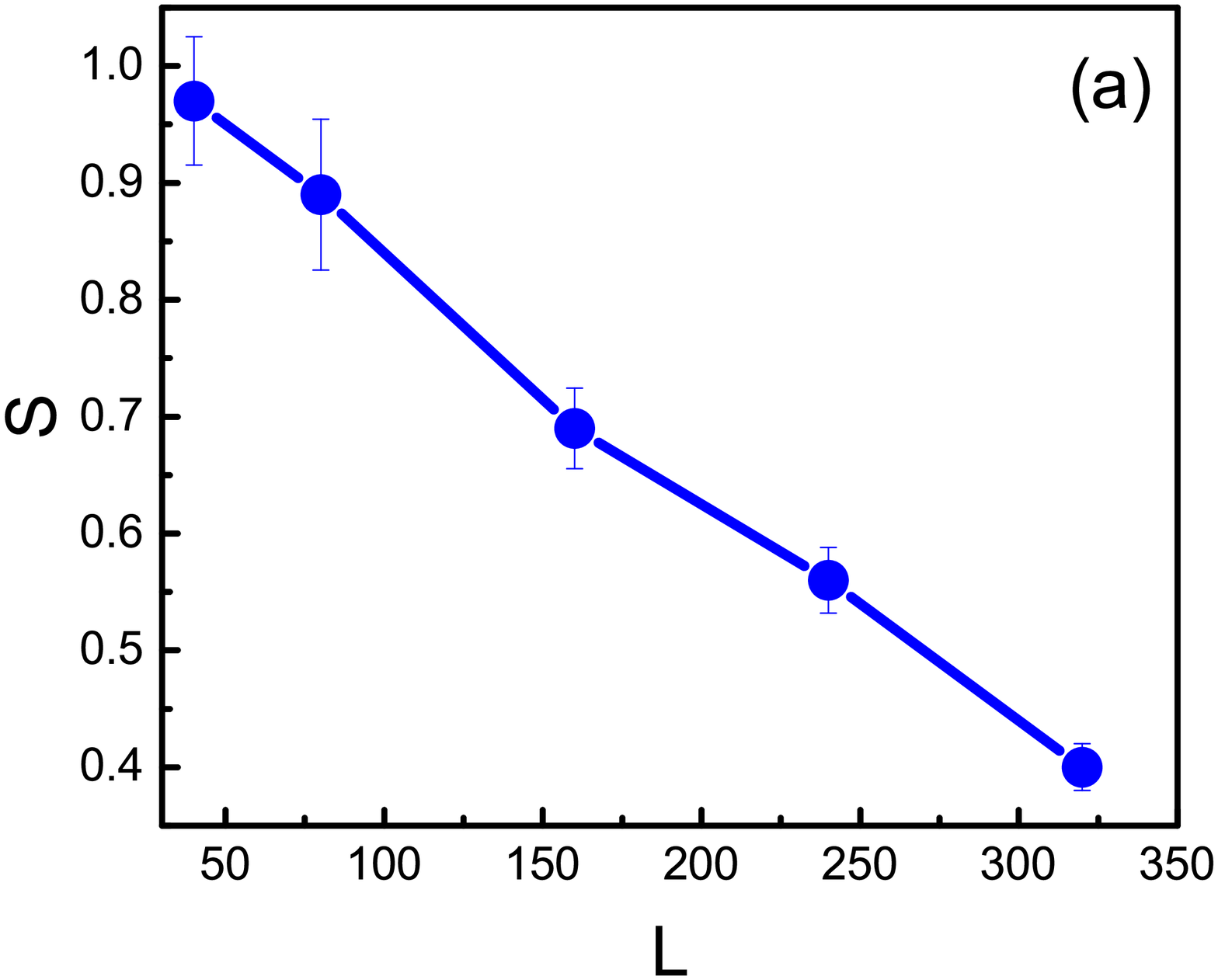}}
\hspace{0.1\linewidth}
\subfigure{\label{fig:subfig:b}
\includegraphics[width=0.4\linewidth]{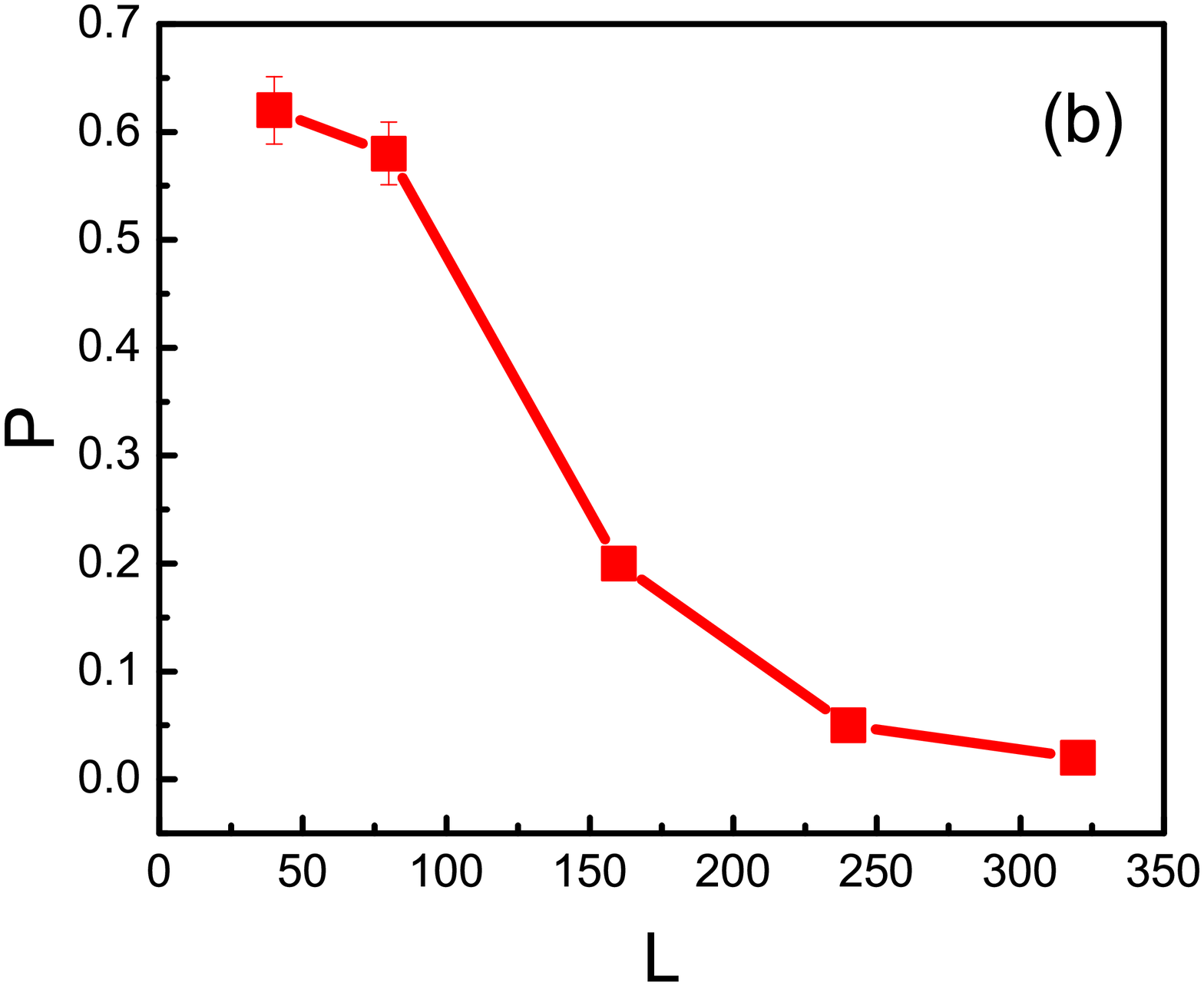}}
\caption{(a) Segregation coefficient $S$ as a function of the box size $L$. (b)Polar order parameter $P$ as a function of a function of the box size $L$. The error bars represent the standard error of the mean. The other parameters are $v_0=1.0$, $D_r=0.001$, $g=0.1$, and $\omega=0.5$.}
\label{fig:subfig}
\end{figure}
\indent Figure 8 shows the box size effects on segregation coefficient and polar order parameter for the demixing case ($g=0.1$ and $\omega=0.5$) at $N=1024$. When the box size $L$ increases from 40 to 320, the  segregation coefficient $S$ decreases from $0.975$ to $0.394$ (from the demixing state to the mixing state) and  the polar order parameter $P$ decreases from $0.62$ to $0.01$.  This can be explained as follows. When $L$ is very large, the the packing fraction is very small and the average distance between particles is large. In this case, when the chirality difference competes with the polar velocity alignment ($g=0.1$ and $\omega=0.5$) particles are demixed in very small sub-regions, however, the two types of particles are still mixed on the whole. In addition, the cluster behaviors are different for different sub-regions, on the whole, particles move in any direction with equal probability, thus $P$ tends to zero for large $L$. For the mixing case (not shown in the figure), both $S$ and $P$ monotonously decreases when the box size $L$ increases.

\begin{figure}
\centering
\subfigure{\label{fig:subfig:a}
\includegraphics[width=0.4\linewidth]{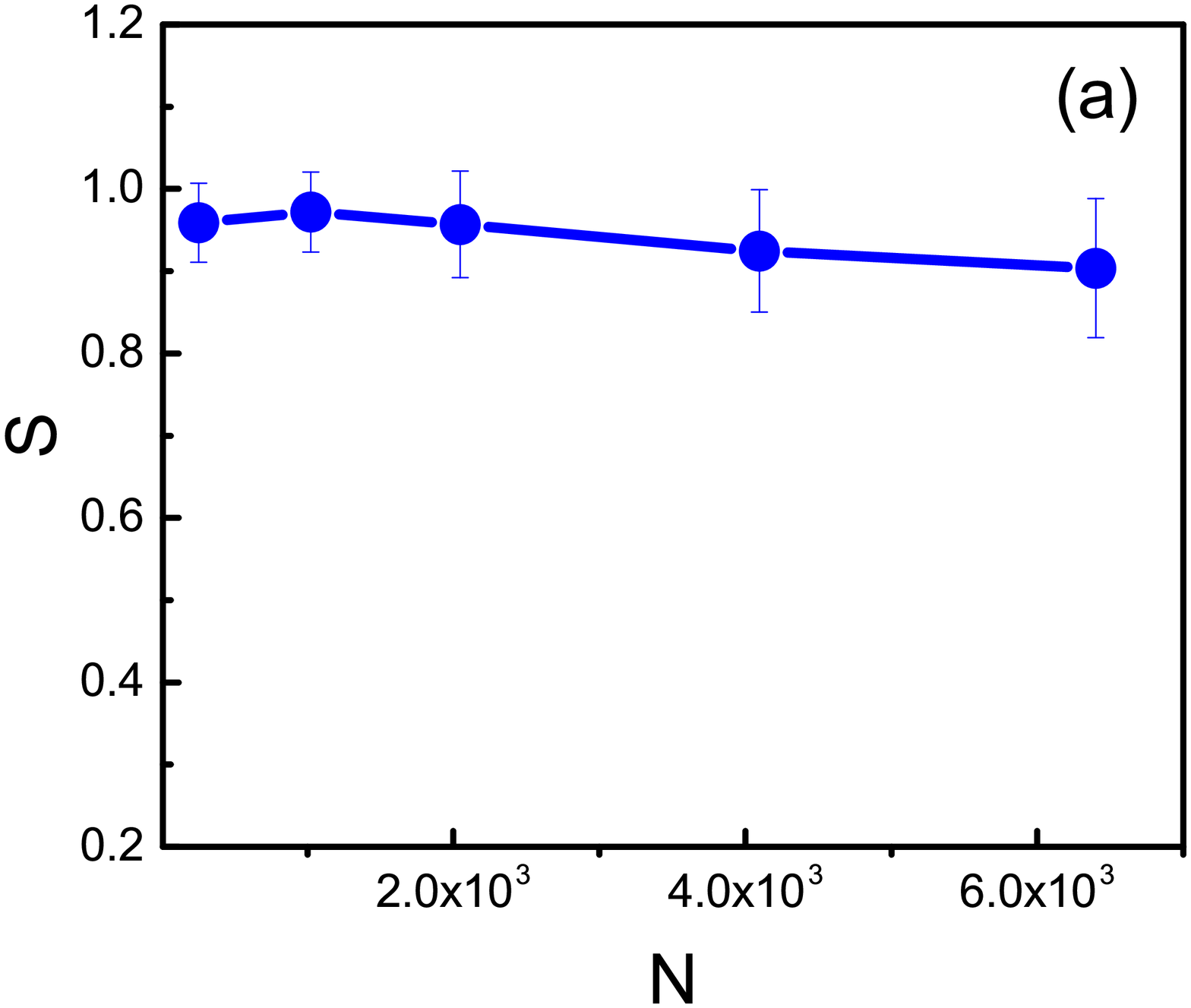}}
\hspace{0.1\linewidth}
\subfigure{\label{fig:subfig:b}
\includegraphics[width=0.4\linewidth]{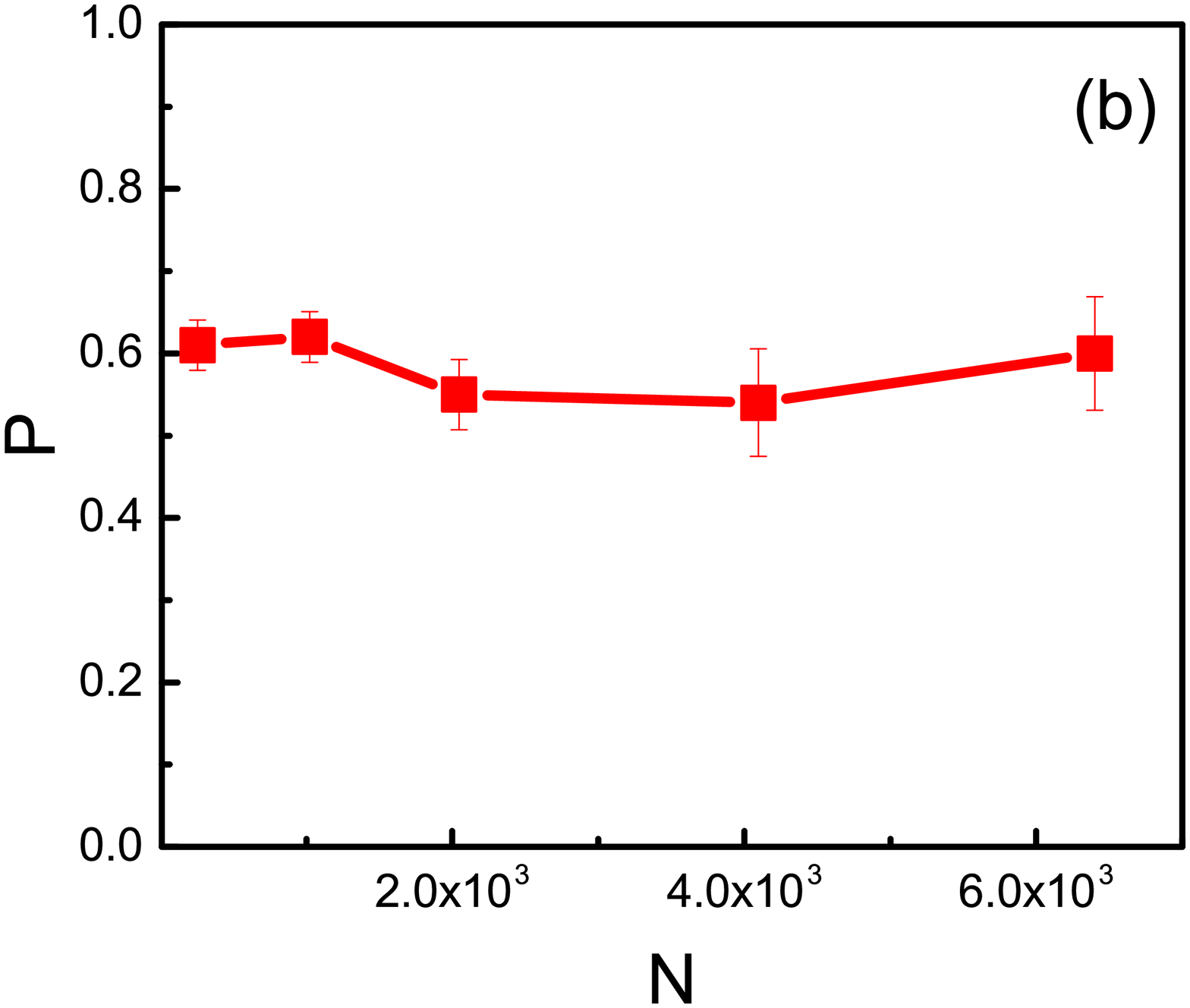}}
\caption{(a) Segregation coefficient $S$ as a function of the particle number $N$ at the fixed packing fraction $\rho=0.50$. (b)Polar order parameter $P$ as a function of a function of the particle number $N$ at the fixed  packing fraction $\rho=0.50$. The error bars represent the standard error of the mean. The other parameters are $v_0=1.0$, $D_r=0.001$, $g=0.1$, and $\omega=0.5$.}
\label{fig:subfig}
\end{figure}

\indent Figure 9 describes the finite effects on segregation coefficient and polar order parameter for the demixing case ($g=0.1$ and $\omega=0.5$) at $\rho=0.50$. When the particle number $N$ is changed at the fixed packing fraction $\rho=0.50$, both segregation coefficient and polar order parameter have changed very little. For all cases shown in Fig. 9, the clockwise particles aggregate in one cluster and the counterclockwise particles aggregate in the other cluster, therefore the two types of particles are demixed. Note that the total integration time will be very long ( much greater than $2\times 10^5$) for large $N$.

\section{Concluding Remarks}
\indent To summarize, the binary mixture of counterclockwise and clockwise particles has been studied in a two-dimensional box with periodic boundary conditions. We have investigated the segregation coefficient and the polar order parameter as functions of the system parameters ($g$, $\omega$,$D_r$, and $v_0$). When the chirality difference competes with the polar velocity alignment, $S>0.8$ and $P\simeq 0.6$, the clockwise particles aggregate in one cluster and the counterclockwise particles aggregate in the other cluster, thus particles are demixed and can be separated. When the chirality difference is dominated,  $S$ is small and $P\to 0$, many small clusters appear and particles are mixed. When the polar velocity alignment is dominated, $S$ is small and $P\to 1$, particles are randomly aggregates in different clusters and completely mixed. When $g$ increases from 0 to $0.5$,  the polar order parameter $P$ increases from $0$ to $1$.  When $\omega$ increases from 0 to $3.0$,  the polar order parameter $P$ decreases from $1$ to $0$. There exists a step in the curves(shown in Figs. 4(b) and 4(d)) which corresponds to particle demixing. Both the segregation coefficient and polar order parameter decrease monotonously with the increase of $D_r$. There exists an optimal value of $v_0$ at which $S$ and $P$ take their maximal values. In addition, both the segregation coefficient and the polar order parameter monotonously decreases when the box size $L$ increases. Both segregation coefficient and polar order parameter have changed very little when the particle number $N$ is changed at the fixed packing fraction $\rho=0.50$.

\indent Our results should be of considerable practical and theoretical interest, because chiral active matter not only encompasses a large class of biological and synthetic micro-swimmers, but also creates a plethora of new phenomena beyond the physics of chiral active matter. Especially, the results we have presented have a potential application in many biological circle swimmers, such as the magnetotactic bacteria in rotating external fields\cite{Cebers}, the bacteria near solid boundaries\cite{Leonardo,DiLuzio}, and the sperm cells with vortex motion\cite{Riedel}.
\section*{Acknowledgements}
\indent This work was supported in part by the National Natural Science Foundation of China (Grants No. 11575064 and No. 11175067), the GDUPS (2016), and the Natural Science Foundation of Guangdong Province (Grant No. 2014A030313426 and No.2017A030313029).

%\bibliography{reference}

\begin{thebibliography}{35}%
\bibitem{Hanggi}P. Hanggi and F. Marchesoni, Rev. Mod. Phys. \textbf{81}, 387 (2009); P. Reimann, Phys. Rep.\textbf{ 361}, 57 (2002).
\bibitem{Bechinger}C. Bechinger, R.D. Leonardo, H. Lowen, C. Reichhardt, G. Volpe, G. Volpe, Rev. Mod. Phys. \textbf{88}, 045006 (2016).
\bibitem{Marchetti0}M. C. Marchetti, J. F. Joanny, S. Ramaswamy, T. B. Liverpool, J. Prost, M. Rao, and R. A. Simha, Rev. Mod. Phys. \textbf{85}, 1143 (2013).
\bibitem{Reichhardt0}C. J. Olson Reichhardt and C. Reichhardt, Annu. Rev. Condens. Matter Phys. \textbf{8}, 51 (2017).
\bibitem{Cates}M. E. Cates, Rep. Prog. Phys. \textbf{75}, 042601(2012).
\bibitem{Reichhardt1x}C. J. Olson Reichhardt and C. Reichhardt, Nature Physics \textbf{13}, 10 (2017).

\bibitem{Volpe}G. Volpe, S. Gigan, and G. Volpe, Am. J. Phys. \textbf{82}, 659 (2014).
\bibitem{McCandlish}S. R. McCandlish, A. Baskarana, and M. F. Hagan, Soft Matter \textbf{8}, 2527(2012).
\bibitem{Stenhammar}J. Stenhammar, R. Wittkowski, D. Marenduzzo, and M. E. Cates, Phys. Rev. Lett. \textbf{114},018301 (2015).
\bibitem{Ma}Z. Ma, Q. Lei, and R. Ni, Soft Matter \textbf{13}, 8940 (2017).
\bibitem{Smrek}J. Smrek and K. Kremer, Phys. Rev. Lett. \textbf{118}, 098002 (2017).
\bibitem{Harder}J. Harder and A. Cacciuto, Phys. Rev. E \textbf{97}, 022603 (2018).

\bibitem{Maggi}C. Maggi, A. Lepore, J. Solari, A. Rizzo, R. Di Leonardo, Soft Matter \textbf{9}, 10885 (2013).
\bibitem{Yang}W. Yang, V. R. Misko, K. Nelissen, M. Kong, and F. M. Peeters, Soft Matter \textbf{8}, 5175 (2012).
\bibitem{Costanzo}A. Costanzo, J. Elgeti, T. Auth, G. Gompper, and M. Ripoll, EPL \textbf{107}, 36003(2014).
\bibitem{Berdakin}I. Berdakin, Y. Jeyaram, V. V. Moshchalkov, L. Venken, S. Dierckx, S. J. Vanderleyden, A. V. Silhanek, C. A. Condat, and V. I. Marconi, Phys. Rev. E \textbf{87}, 052702(2013).
\bibitem{Nourhani}A. Nourhani, V. H. Crespi, and P. E. Lammert, Phys. Rev. Lett. \textbf{115}, 118101 (2015).
\bibitem{Weber}S. N. Weber, C. A. Weber, and E. Frey, Phys. Rev. Lett. \textbf{116}, 058301 (2016).
\bibitem{Kumari}S. Kumari, A. S. Nunes, N. A. M. Araujo, and M. M. T. Gama, J. Chem. Phys. \textbf{147}, 174702 (2017).


\bibitem{Mijalkov}M. Mijalkov and G. Volpe, Soft Matter \textbf{9}, 6376 (2013).
\bibitem{Reichhardt}C. Reichhardt and C. J. Olson Reichhardt, Phys. Rev. E \textbf{88}, 042306 (2013).
\bibitem{Scholz}C. Scholz, M. Engel, and T. P$\ddot{o}$schel, Nat. Commun. \textbf{9}, 931 (2018).


\bibitem{Nguyen}N.H. P. Nguyen, D. Klotsa, M. Engel, and S. C. Glotzer, Phys. Rev. Lett. \textbf{112}, 075701 (2014).


\bibitem{Dolai}P. Dolai, A. Simha, and S. Mishra, arXiv:1706.02968 (2017).
\bibitem{Agrawal}A. Agrawal and S. B. Babu, arXiv:1709.09085 (2017).
\bibitem{Wysocki}A. Wysocki, R. G. Winkler, and G. Gompper, New J. Phys. \textbf{18}, 12303 (2016).

\bibitem{Ai1}B. Q. Ai, Y. F. He, and W. R. Zhong, Soft Matter \textbf{11}, 3852 (2015).
\bibitem{Ai2}Q. Chen and B. Q. Ai, J. Chem. Phys. \textbf{143}, 104113 (2015).
\bibitem{Ai3}B. Q. Ai, Sci. Rep. \textbf{6}, 18740 (2016).

\bibitem{Shin}J. Shin, A. G. Cherstvy, and R. Metzler, New J. Phys. \textbf{16}, 053047 (2014).

\bibitem{Leonardo}R. Di Leonardo, D. DellArciprete, L. Angelani, and V. Iebba, Phys. Rev. Lett. \textbf{106}, 038101 (2011).
\bibitem{DiLuzio}W. R. DiLuzio, L. Turner, M. Mayer, P. Garstecki, D. B. Weibel, H. C. Berg, and G. M. Whitesides, Nature (London) \textbf{435}, 1271 (2005).
\bibitem{Cebers}A. Cebers, J. Magn. Magn. Mater. \textbf{323}, 279 (2011).
\bibitem{Riedel}I. H. Riedel, K. Kruse, and J. Howard, Science \textbf{309}, 300 (2005).


\bibitem{Vicsek}T. Vicsek and A. Zafeiris, Phys. Rep. \textbf{517}, 71 (2012).
\bibitem{Levis}D. Levis and B. Liebchen, J. Phys.: Condens. Matter \textbf{30}, 084001 (2018).
\bibitem{Liechen}B. Liebchen and D. Levis, Phys. Rev. Lett. \textbf{119}, 058002 (2017).
\bibitem{Mart¨ªn-Gomez}A. Mart¨ªn-Gomez, D. Levis, A. Diaz-Guilera, and I. Pagonabarraga, arXiv:1801.01002 (2018).

\bibitem{Yang1}X. Yang, M. Manning, and M. C. Marchetti, Soft Matter \textbf{10}, 6477 (2014).

\bibitem{Michael}M. E. Cates and J. Tailleur,  Annu. Rev. Condens. Matter Phys. \textbf{6}, 219(2015).

\end{thebibliography}
%\end{document}

%

\end{document}